\documentclass[a4paper,12pt]{article}
\usepackage[utf8]{inputenc}
\usepackage{tikz}
\usepackage{amsmath}
\usepackage{mathtools}
\usepackage{bm}
\usepackage{algorithm}
\usepackage{algpseudocode}
\usepackage{graphicx}
\usepackage[font={small}]{caption}
\usepackage{subfigure}
\usepackage[margin=1in]{geometry}
\usepackage{hyperref}
\usepackage{natbib}
\usepackage{array}
\usepackage{authblk}
\usepackage{color}
\usepackage{placeins} 
\usepackage{dcolumn}
\usepackage{lscape}
\usepackage{longtable}
\usepackage[official]{eurosym}

\newcolumntype{L}[1]{>{\raggedright\let\newline\\\arraybackslash\hspace{0pt}}m{#1}}
\newcolumntype{C}[1]{>{\centering\let\newline\\\arraybackslash\hspace{0pt}}m{#1}}
\newcolumntype{R}[1]{>{\raggedleft\let\newline\\\arraybackslash\hspace{0pt}}m{#1}}

\begin{document}

\title{What do online listings tell us \\ about the housing market?}
\author[1]{Michele Loberto}
\author[1]{Andrea Luciani}
\author[2]{Marco Pangallo}
\affil[1]{Department of Economics and Statistics, 

Banca d'Italia, Rome 00184, Italy }
\affil[2]{Institute of Economics and Department EMbeDS, 

Sant'Anna School of Advanced Studies, Pisa 56127, Italy}

\date{\today}
\maketitle

\begin{abstract}
Traditional data sources for the analysis of housing markets show several limitations, that recently started to be overcome using data coming from housing sales advertisements (ads) websites. In this paper, using a large dataset of ads in Italy, we provide the first comprehensive analysis of the problems and potential of these data. The main problem is that multiple ads (``duplicates'') can correspond to the same housing unit. We show that this issue is mainly caused by sellers' attempt to increase visibility of their listings. Duplicates lead to misrepresentation of the volume and composition of housing supply, but this bias can be corrected by identifying duplicates with machine learning tools. We then focus on the potential of these data. We show that the timeliness, granularity, and online nature of these data allow monitoring of housing demand, supply and liquidity, and that the (asking) prices posted on the website can be more informative than transaction prices.
\end{abstract}

\let\oldthefootnote\thefootnote
\renewcommand{\thefootnote}{\fnsymbol{footnote}}
\footnotetext[1]{Email: \href{mailto:michele.loberto@bancaditalia.it}{michele.loberto@bancaditalia.it}, \href{mailto:andrea.luciani@bancaditalia.it}{andrea.luciani@bancaditalia.it}, \href{mailto:marco.pangallo@maths.ox.ac.uk}{marco.pangallo@santannapisa.it}. The views expressed in this paper are those of the authors and do not reflect the views of Banca d'Italia. We are extremely grateful to Immobiliare.it for providing the data and for their assistance. We thank all the seminar and workshop participants at Banca d'Italia, Bank of England, Danmarks Nationalbank, Eurostat, JRC Centre for Advanced Studies (JRC-CAS), for useful comments and discussions. Special thanks go to Attilio Mattiocco, Juri Marcucci, Diana Nicoletti for their assistance and feedback in all the phases of this project. An earlier version of this paper circulated under the title ``The potential of big housing data: an application to the Italian real-estate market''; Marco Pangallo performed most of his work on this project while he was affiliated to the Mathematical Institute and Institute for New Economic Thinking at the University of Oxford. All mistakes are our own.}
\let\thefootnote\oldthefootnote


\newpage

\section{Introduction}
\label{sec:intro}

\indent The lack of comprehensive data on housing transactions has so far represented a limit to our understanding of the housing market. Microdata on home sales are available to researchers only in a few countries and, unfortunately, many of these sources show limitations in the spatial or in the temporal dimension. Moreover, most of the time microdata on home sales come from administrative datasets, for which information is not collected for statistical purposes, but to enforce tax payments and property rights.\footnote{One of the most popular administrative datasets on home sales is the Multiple Listing Service (MLS) database for the US housing market. However, most studies using MLS focused on a few Metropolitan Statistical Areas only, due to the lack of a centralized MLS. In Italy, administrative microdata on housing transactions are not available for research because of privacy concerns. The UK is one of the few countries where microdata are timely available, but those data contain poor information about physical characteristics of homes.} Even more challenging is finding comprehensive information about homes and the full history of housing transactions, from the time a dwelling enters the market up the actual transaction.\footnote{ \cite{merlo2004bargaining} provide one of the most complete analyses on the full history of housing transactions. However, because their data are hand-collected from real estate agents, their sample size is extremely limited: 780 housing units.  } 

More recently, online data from marketplace websites (such as Zillow or Trulia in the US, or Zoopla in the UK) started filling some gaps. From these sources we can retrieve detailed information about characteristics of each housing unit and their precise location. We know how long homes have been listed on the market, and we have a broad picture of housing supply at each point in time. Most importantly, online listing data provide additional information useful to analyze the housing market, such as buyers' search behavior \citep{piazzesi_schneider_stroebel_2015}. Finally, online listings are available in real time, allowing a more timely estimation of house price indexes \citep{anenberg_laufer_2017}.

Although the potential of online listings data for policy and research is enormous, across all fields of economics, several concerns about data quality arise. First, online listings may well fail to provide universal coverage, and so lead to non-representative results. Second, these data are less structured and controlled than administrative data, and there might be hidden factors influencing the generation of the data. Finally, online data could have other sorts of measurement errors, mostly if compiled by hand. 

In this paper we present new evidence about the potential of online home listing data and the trade offs that analysts may face when using these data. Some drawbacks are already well-known: listing data provide no information about sale prices and about what caused a delisting, i.e. if a home has been sold or withdrawn. Our first contribution is to show that there can be other limitations in the quality of these data that may introduce severe measurement issues. Up to our knowledge, this is the first work that performs a comprehensive analysis of the reliability and representativeness of online home listings. After discussing drawbacks, our second contribution is to discuss some novel applications of these data, highlighting how they can complement traditional statistical sources.  

We analyze a database containing housing sales advertisements (ads) published on Immobiliare.it, the most popular online portal for real-estate services in Italy. The data cover the period from January 2015 up to December 2018 and include sales offers of residential units in 109 cities. We observe the history of about 1.4 millions ads at a weekly frequency, from the time they were created up to the time they were removed from the website. Overall, the database contains almost 28 million records.

Although this work is based on Italian data only, the same issues encountered in our analysis affect online listings in other countries, as most of the home listings websites across countries are very similar. Marketplace websites provide a platform where sellers or brokers can post their offers, with no strict control on what is published. In the case of the housing market, this is problematic because multiple brokers can intermediate the same good. The owner of a home can entrust several real estate agencies for the sale of her dwelling and each of these agencies could publish a different ad. Thus, there can be multiple ads referring to the same dwelling. The drawback of duplicate ads is that they can make online listings non-representative of the housing supply, especially if the number of ads associated to each dwelling is not randomly distributed. As a consequence, the lack of strict control of what is published calls for cautiousness in the use of online listing data, especially in countries where open mandates to sell with multiple agencies are a relevant feature of the housing market.\footnote{ Usually, the main source of profits for home listings websites are the fees paid by the users to publish their ads. For this reason, these website have no incentive to check if multiple ads refer to the same dwelling. } 

To quantitatively assess the measurement error associated with the use of these data, we employ machine learning algorithms to identify and remove duplicate ads, so as to group all listings referring to the same home. We resort to machine learning tools because we have found that using simple heuristic rules based on the location and similarity in physical characteristics provides imprecise classification. The share of listed homes that have been associated to one unique ad for all the listing period is 77 percent. Only 15 percent of listed homes have been brokered by more than one real estate agent, but these homes represent 35 percent of all ads.

After removing duplicate ads, we show that online listings provide information consistent with existing statistical sources. Duplicates introduce a severe measurement error, especially for local housing markets. Moreover, duplicates are not randomly distributed among listed homes. The propensity to post multiple ads increases when the asking price is relatively high and, more generally, home sellers face weak demand for their homes. By posting multiple ads, home sellers can immediately increase the visibility of their homes, although this effect is only temporary. Home sellers and brokers increase their effort in advertising when homes are relatively unattractive. 

Once we have shown how it is possible to increase the reliability of online listing data, we focus on their potential for the analysis of the housing market. We consider three applications, namely how online listings can be used to assess the evolution of demand, supply and house prices. 

As a proxy of demand, we consider the number of visits to the webpage of each ad during each week.\footnote{This proxy is complementary to web searches, which have for example been used recently by \cite{piazzesi_schneider_stroebel_2015}.} When individuals visit the webpage of an ad, they convey information about the characteristics, the location and the price of the home they are searching for. By aggregating all this information, we can understand which area households are searching more intensely and what they are looking for.
 We provide evidence that online interest indeed proxies housing demand. At micro (dwelling) level, high online interest predicts a lower time on market and a lower probability that a downward revision of the asking price occurs. By aggregating the number of page views we can build a measure of market tightness: we show that this indicator is a good predictor of market liquidity and house prices, as suggested by the recent literature \citep{carrillo_etal_2015, wu_brynjolfsson_2015, vandijk_francke_2017}.\footnote{We presented the results about demand in a previous publication \citep{pangallo_loberto_2018}. We still find it useful to report them here, both because that publication was directed at a very different audience, and because the results reported in this paper are valid for a time span that is twice as long, highlighting robustness of our findings. }

Online listings also make it possible to reconsider the standard definition of housing supply. This is usually defined as the total stock of homes \citep{glaeser_gyourko_2018}, and according to this definition a change in housing supply is only due to new constructions. While this definition may be most appropriate for long-run analyses, we think that it is misleading in the short-medium run and would hide relevant housing market facts. For example, we show that households take market conditions into account before deciding whether to put their home up for sale. In particular, the quality of listed existing homes, as proxied by some physical characteristics, improves with better market conditions. This is consistent with \cite{ngai_sheedy_2018}, who show that household decision to move house is affected by general business cycle conditions, and house sales are very correlated with listings.

Finally, we look at the potential of listings data for the assessment of the dynamics of house prices. Home listings websites usually provide asking prices only.\footnote{In a few countries such as Netherlands and Denmarks, both asking and transaction prices can be available because the online portals are owned by associations of real estate agencies.} Asking prices can be used to nowcast and forecast sale prices, but a sound estimate of the average discount to the initial asking price is needed, because the average discount to the initial asking price changes over time with market conditions. However, forecasting sale prices may not be the only goal. For example, we show that asking prices may be a more robust indicator of the timing of the house price cycle than sale prices. Indeed, sale prices are more volatile in the short run, mostly because of the relative small number of sales compared to listings, and this hinders their ability to identify turning points in real time. At the opposite, we provide evidence that asking prices are a more suitable indicator for this task, since they are more stable and more timely.

This paper is organized as follows. Section \ref{sec:data} describes the Immobiliare.it ads dataset, introducing the issue of duplicate ads. In Section \ref{sec:deduplication_text} we describe how we create the final dataset of housing units, and we validate our dataset by comparing it with the official statistical sources available. In Section \ref{sec:comp_raw_dedupl} we illustrate the distortions introduced by duplicate ads.  Finally, in Section \ref{sec:measure_stats} we show how online listing data can be used to measure demand, supply and house prices. Section \ref{sec:conclusions} concludes.

\section{Data}
\label{sec:data}

We analyze a dataset of home listings published on Immobiliare.it, the largest online portal for real-estate services in Italy. This dataset covers the full country. Because small towns and villages may have representativeness issues, we only consider listings in the 109 main cities that are capitals of the NUTS-3 Italian regions. About 18 million people live in these cities, and the number of home sales is about one-third of all transactions in Italy.

Immobiliare.it provides us with weekly snapshots of all ads visible on their website every Monday, since 2016, January 4 until 2018, December 31. For 2015 only quarterly snapshots are available.\footnote{Data are available for the following days in 2015: January 5, April 25, September 7, December 28.} 

For each ad we have detailed information about physical characteristics and exact location of the dwelling (see Appendix \ref{sec:app_description_data} for the complete list of variables). We keep track of all variations concerning asking prices and the number of times that the webpage of the ad has been visited (\textit{clicks}). We also know the date when the ad was created and the date when it was removed. Unfortunately, as usual with listing data, we do not know if an ad has been removed because the home was sold or withdrawn from the market.

The dataset counts 1,402,798 ads. Since we observe ads at weekly frequency, the total number of records is almost 28 millions. Most ads remain unchanged between two weekly snapshots, and the average turnover is about 5\%. Real estate agents post about 92\% of the ads, the others are posted by households or construction firms.

We divide the territory of each city into local housing markets using the partition developed by OMI ("microzones"),\footnote{This partition is periodically revised to satisfy these criteria and to better approximate local housing markets. The last revision dates back to 2014.} a branch of the Italian Tax Office. The elements of this partition are contiguous areas of the city territory that satisfy strict requirements in terms of homogeneity of house prices, urban characteristics, socioeconomic characteristics, as well as endowment of services and urban infrastructures. Thus, unlike census tracts, these microzones can be considered as ``local housing markets''. For each of these microzones, OMI provides estimates of the minimum and maximum house price per square meter on a six-month basis. Table \ref{tab:descriptive_statistics_zoneomi} in Appendix \ref{sec:addtabfig} reports some descriptive statistics about these local housing markets.  

Finally, we make use of information coming from the Italian Housing Market Survey, a quarterly survey that covers a large sample of real estate agents. A detailed discussion about all data sources can be found in Appendix \ref{sec:app_description_data}.

\subsection{Duplicate listings}
\label{sec:duplicates}

The main concern with marketplace websites -- such as Immobiliare.it, Craigslist, Zoopla, ImmobilienScout24 and many others -- is the absence of strict control over the ads published by the users. These websites are market platforms that allow home sellers and brokers to advertise the sale of a housing unit in exchange for a fee. Rigorous checks on ads published by users are costly or even unfeasible. As a consequence, online listings may not be reliable for economic analyses. 

In this paper, we assess the implications of a specific source of measurement error. Online listings can contain a large amount of duplicate ads, i.e. multiple ads can be associated to the same dwelling. This may be due to various reasons. First, in some countries owners sell their home through open listing agreements. Under these agreements, two or more real estate agents are entitled to sell the same dwelling, and each of them could publish a different ad. Second, a real estate agency could post multiple ads for the same home.\footnote{ This is a trick that real estate agencies can use to misrepresent the time on market of the listing, and make a listing look like new. They may want to do that because potential buyers can order the result of their search on the website from the most recently published ad to the oldest. Moreover, posting a new ad provides greater visibility to the listing, because many potential buyers receive notifications about new listings through the email-alert service.} Third, a mandate to sell expires, and the home seller signs a listing agreement with a new agent. If the old agent does not immediately delete the ad, and the new agent posts a new one, two ads for the same dwelling exist at the same time. Even if this does not happen and the two ads are not simultaneously visible on the website, we still need to know that these ads refer to the same dwelling. 

For a quantitative assessment of this phenomenon, for the three-year period 2016-2018 we compare the number of delistings that we observe in our dataset with the number of home sales, provided by OMI. As can be seen in Table \ref{table:ads_tomsales}, the number of home sales is about 60 percent of the number of delistings, with large volatility across different cities (40\% in Florence, 70\% in Naples). Let us assume that each ad is associated to a different dwelling. In this case, the share of listings ending with a sale should be lower than 60 percent. This estimate is broadly consistent with studies on the US housing market \citep{anenberg_laufer_2017,carrillo_williams_2019} but definitely small compared to evidence from the UK \citep{merlo2004bargaining}.\footnote{ According to \cite{anenberg_laufer_2017} and \cite{carrillo_williams_2019}, in the US about half of the delistings results in withdraws. In a sample of listings from the UK, \cite{merlo2004bargaining} find that withdraws are about 25 percent of the delistings. }  Since our dataset is mostly representative of home sales brokered by estate agents -- that do not represent all transactions -- the assumption that each ad is associated to a different dwelling would imply that the share of sales over all delistings could be even lower than 50 percent. We believe that such a share of sales would be excessively low for the Italian market.

\begin{table}[ht]
\centering
\footnotesize
\begin{tabular}{ccccc}
  \hline
  \hline
Year & Delistings & Sales & \multicolumn{2}{c}{Time on market} \\
\cline{4-5} 
 &  &  & Listings & Survey \\
  \hline
2016 & 335,181 & 178,690 & 5.1 & 7.5 \\ 
  2017 & 312,584 & 186,657 & 4.9 & 6.3 \\ 
  2018 & 321,840 & 197,506 & 4.4 & 6.6 \\ 
   \hline
   \hline
\end{tabular}
\caption{Number of delistings, house sales and time on market (months). Data on sales and time on market come from the Immobiliare.it dataset and from the OMI and Italian Housing Market Survey (see Appendix \ref{sec:app_description_data}).} 
\label{table:ads_tomsales}
\end{table}

In addition to the potentially low share of sales, the average time on market computed on listings -- as the number of months between the initial listing and the delisting  -- is significantly lower than the estimates provided by real estate agents in the Italian Housing Market Survey (Table \ref{table:ads_tomsales}). Using listings implies an underestimation of about two months for the time on market.

A plausible explanation for the potentially low share of sales and the underestimation of the time on market is that there is not a one-to-one correspondence between ads and housing units actually put on the market. Because of that, individual ads do not always provide information about the full history of a home listing, and multiple ads in the dataset can refer to the same home.  

The presence of duplicate ads has two implications. First, ads provide a biased representation of the supply, especially at granular levels. Our major concern is that the presence of duplicates is not random but associated with the physical characteristics of the house, the urgency of the owner to sell soon or difficulties to find a buyer. Second, the sale of a single dwelling can be associated with several ads posted in different periods, possibly from different intermediaries. If this second problem occurs, we cannot observe the full history of  housing sales.  

More rigorous checks by website administrators about information published on the market platform cannot solve all issues associated with online listing data, especially in countries where open listing agreements are a relevant feature of the housing market. Thus, there will always be some concern that online listings are not representative of the actual housing market dynamics. A first contribution of this paper is to quantitatively assess the magnitude of the resulting potential distortion.

\section{Identification of home listings}
\label{sec:deduplication_text}

We identify duplicate ads using machine learning tools. In this way, we depart from the original dataset of ads and build a new dataset of listings, in which the unit of observation is a home instead of an ad. 

We use machine learning tools because, most of the times, we cannot identify an exact matching between characteristics of the houses reported in two duplicate ads. In these situations, heuristic rules imposed by the analyst fail, while machine learning algorithms autonomously learn the best criteria that identify duplicates provided the training sample is sufficiently large. After identifying duplicates, we combine them as if they were a single ad. 

Here we sketch the steps of the algorithm, while the details are left to Appendix \ref{sec:deduplication} and \cite{loberto_etal_18}. Our approach is based on standard methodologies adopted for data deduplication \citep [see][]{naumann_herschel_2010,christen_2012}, that we adjust to better tackle the specifics of our dataset. The deduplication process consists of three steps.
\bigskip

\noindent \textbf{Data preparation}. To identify if two ads refer to the same dwelling, we have to compare locations and characteristics of the houses described in the ads. This operation is complicated by the fact that the geographical coordinates or the address may not be precise enough and some information is not objective, but based on the best judgment of the home seller/broker.\footnote{The seller/broker of the home can identify the location on a map or, alternatively, provide the address as an input. The fact that two different tools are available -- and the lack of precision of the user -- gives rise to the possibility that the same dwelling has a slightly different geolocation in different ads. That is not an issue in rural areas, but it is in urban areas with a high concentration of housing units. } 

For this reason, we cannot look for a perfect matching between home characteristics and have to build measures of partial similarity. To do that, we consider some ordered categorical variables -- such as maintenance status, for which the various categories can be assigned an order -- as numerical to meaningfully calculate the degree of similarity between these characteristics. Overall, for numerical and ordered categorical variables, we compute the distance between characteristics reported in the two ads. Only for binary and categorical variables that are not ordered we look for a perfect matching between the two ads. 

The textual description of the home provided in the ad performs a dual role. First, by using semantic analysis, information extracted from the textual description allows to impute missing data. That is important because the best way to identify duplicates is to retrieve from the ads as much information as possible. Second, we use the textual description as a further variable to identify if two ads refer to the same dwelling. We use the Paragraph Vector (or \textit{doc2vec}) algorithm proposed by \cite{le_mikolov_2014}, in which a neural network learns about the order and semantics of words, to associate a numeric vector to the textual description. In this way, the ``distance'' between two descriptions can be measured by computing the cosine distance between the associated vectors. 
\bigskip

\noindent \textbf{Classification}. We check if two ads are duplicates only if they are sufficiently close, both in terms of geographical distance and listing price. For each of these pairs we have to decide if the ads refer to the same housing unit. To do so, we compare the characteristics of the dwellings described in the ads, and based on some rules we classify them as \textit{duplicates} or \textit{not duplicates}.  

To identify these rules we use a machine learning algorithm, the C5.0 classification tree proposed by \cite{Quinlan1993c45}. This algorithm autonomously learns which variables are most relevant to identify duplicates, once it is supplied with a sufficiently large \textit{training sample}, i.e. a dataset of pairs of ads of which we know with certainty if they are duplicates or not. We manually constructed the training sample by looking at pairs of ads on the website, and deciding whether the two ads refer to the same housing unit based on the pictures of the two dwellings. 

We use different classification trees, depending on whether the two ads were posted by the same user or by different users.\footnote{ We observed that the degree of similarity between two duplicate ads is lower when they are posted by two different users instead of the same user. } Each training sample includes about 9000 pairs of ads, and both classification trees show a high degree of accuracy (see Table \ref{table:results_models} in  Appendix \ref{sec:deduplication}).   

Finally, we use the classification tree for prediction. The algorithm outputs a probability that the two ads are duplicates. If this probability is larger than 0.5, we consider the two ads as referring to the same housing unit. 
\bigskip

\noindent \textbf{Clusterization}. The output of the previous step is a list of pairs of duplicate ads. Since multiple pairs can refer to the same dwelling, we have to create clusters of all ads referring to the same home. To do so, we use methods from graph theory and consider a cluster of ads as referring to the same housing unit if an internal similarity condition for the cluster is satisfied. Finally, we aggregate information coming from different ads by computing for each variable the average or the most common feature observed across ads.  
\bigskip

\noindent \textbf{Final dataset}. The number of houses -- or ``true'' listings -- is only 67 percent of the number of ads (about 950 thousand housing units). Looking at the distribution of houses per number of associated ads, we find that duplicates are concentrated over a small share of houses: about 77 percent of dwellings only have one associated ad, 13 percent have two duplicate ads, and 10 percent have more than two duplicates. 

Open listing agreements with many agents seem to be a main source of duplicate ads. To see that, consider that only 15 percent of homes were listed with more than one agency, but these homes account for 35 percent of the ads. Conversely, 7 percent of homes have multiple ads that were posted by the same broker. These homes account for 13 percent of the ads.  

Considering a single daily snapshot, the number of listed houses is on average 87 percent of the number of ads, meaning that by taking a snapshot of the data on any specific day we expect that only 13 percent of the ads are duplicates. These figures are consistent with those concerning the full sample because duplicate ads for the same listed home grow over time: new ads are created while old ads are deleted, and that gives rise to a huge number of delistings and new listings. We find confirming evidence when we consider only homes with multiple corresponding ads. Every week, for 90 percent of them, at most two duplicate ads are on average visible. This figure can be compared with the share of homes with two ads out of homes with multiple ads in the full sample, which is $10/(10+13)=57\%$.

Finally, the share of duplicates over total ads increases with city size, and there is significant variability across cities. For example, the ratio between the number of ads and housing units is equal to 1.4 for Naples and 1.8 for Rome. Therefore, an additional implication of duplicates is that they can make the comparability across cities difficult.

\subsection{Validation}
\label{sec:validation}

To validate the quality of our deduplication procedure we compare information coming from the final dataset with other well-established statistical sources. 

\begin{figure}[h]
\centering
\subfigure[Sales] 
{\includegraphics[width=0.45\linewidth]{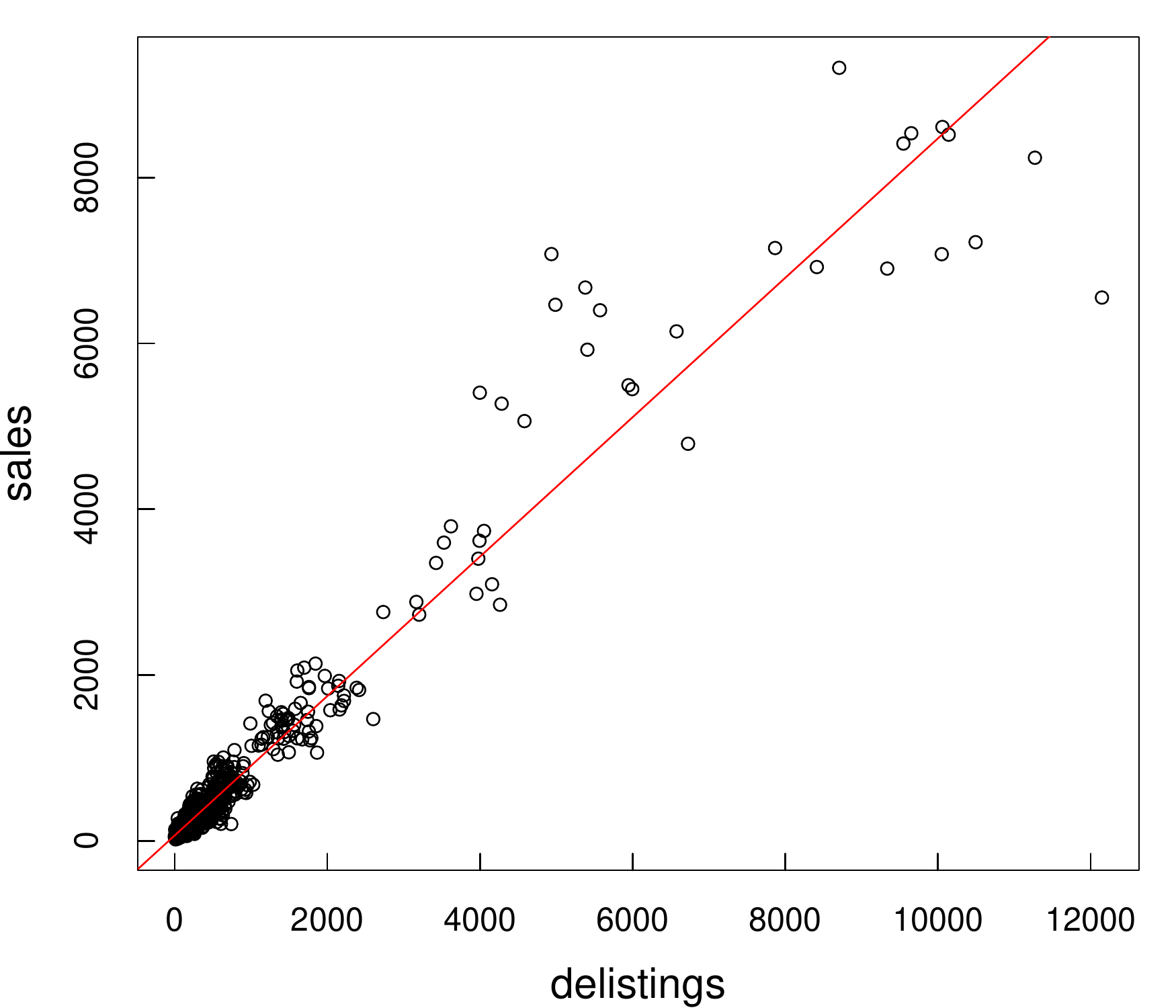}  \label{fig:dwell_transactions_full}}
 \hspace{0mm}
\subfigure[Prices] 
{\includegraphics[width=0.45\linewidth]{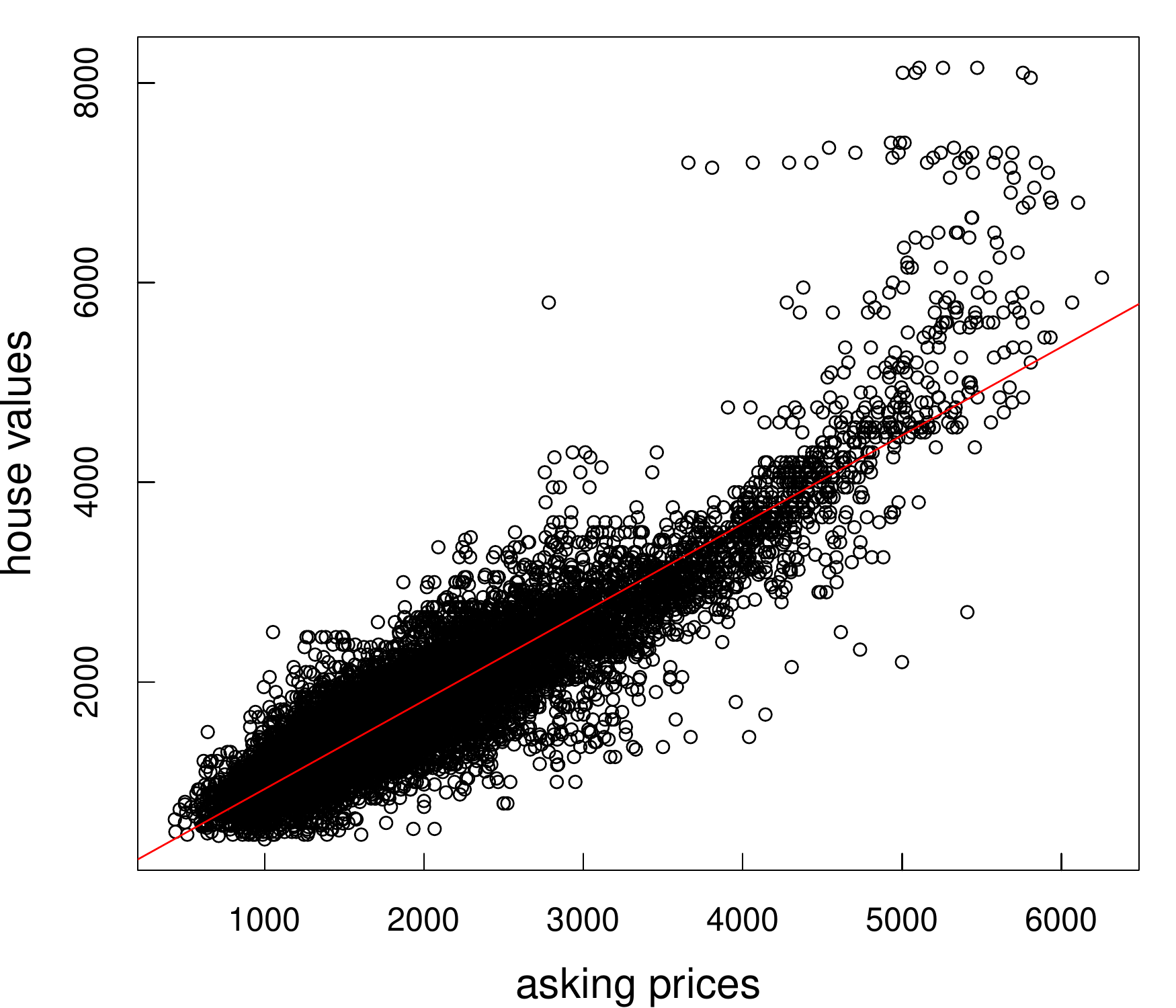} \label{fig:dwell_priceszonaomi}}
\caption{Sales and prices. Listing data are shown on the horizontal axis while data coming from official statistical sources are shown on the vertical axis.}
\end{figure}

First, we compute the number of delistings and home sales in each city (obtained from OMI) at quarterly frequency and we find that these two variables are strongly correlated (Figure \ref{fig:dwell_transactions_full}): their correlation coefficient is 0.94. Notice that a delisting now represents the effective exit of a home from the market instead of the removal of an ad from the website. Table \ref{table:ads_tomsales2} compares the absolute number of delistings and home sales. Compared to Table \ref{table:ads_tomsales},  the numbers seem more plausible once we take into account that not all houses sold during these years have been listed on Immobiliare.it. 

\begin{table}[ht]
\centering
\footnotesize
\begin{tabular}{ccccc}
  \hline
  \hline
Year & Delistings & Sales & \multicolumn{2}{c}{Time on market} \\
\cline{4-5} 
 &  &  & Listings & Survey \\
 \hline
2016  & 207,120 & 178,690  & 6.7 & 7.5 \\ 
  2017  & 187,443 & 186,657  & 6.7 & 6.3 \\ 
  2018  & 189,505 & 197,506  & 6.3 & 6.6 \\ 
   \hline
   \hline
\end{tabular}
\caption{Number of delistings, house sales and time on market (months)} 
\label{table:ads_tomsales2}
\end{table}

We find a robust correlation with official data also when we consider prices (the correlation is 0.82; Figure \ref{fig:dwell_priceszonaomi}). In this case our results are even stronger, because we have official estimates from OMI for each local housing market, so we can compare listing prices and average home values per square meter at a finer granularity. The non-linearity observed for very high home values is probably due to the fact that OMI estimates refer to the average value of all houses located within the local market, while the most expensive and prestigious houses are likely to be less liquid and therefore less represented among listed houses. 

Moreover, for each local housing market we compute the ratio between listing prices and actual home values per square meter. On average, we find that during the three years 2016-2018 the discount on asking prices was about 12\%, a value consistent with the evidence provided by the Italian Housing Market Survey. 

Finally, we look at time on the market. After our deduplication procedure, listings provide an estimate of the time on market overall consistent with the Italian Housing Market Survey (see Table \ref{table:ads_tomsales2}). We find a significant deviation only for 2016 when listings underestimate time on market. That is plausible because that is the first year for which we have weekly data. Some of the houses listed in 2016 may have been initially listed in 2015, but since  for 2015 we only observe quarterly snapshots, it is possible that we were not able to reconstruct the full history of these listings due to difficulties in identifying duplicates.  

Overall, information coming from our final dataset of listings is consistent with official statistical sources. We consider this as evidence of the efficacy of our deduplication procedure.

\section{The implications of duplicate listings}
\label{sec:comp_raw_dedupl}

The first contribution of this paper is to assess what kind of distortions are introduced by duplicate ads. First, we quantify the measurement error introduced by duplicate ads. We document that it is quite small when the level of data aggregation is sufficiently high, but more serious when studying more disaggregated housing markets. Second, we show that presence of multiple ads related to the same dwelling is not random, and home sellers or brokers post multiple ads to attract more attention. 

\subsection{Measurement error}

To quantify the measurement error introduced by duplicate ads we compare some statistics calculated on the original dataset of ads and the final dataset of listings. We check if the two dataset provide similar information about delistings and asking prices. We make this comparison both at the city and local housing market level, as the measurement error possibly has different magnitude depending on the level of granularity that is considered. Average prices per square meter and delistings are averaged over quarters, obtaining quarterly data spanning from 2016Q1 to 2018Q4. 

\begin{figure}[h]
\centering
\subfigure[Delistings] 
{\includegraphics[width=0.45\linewidth]{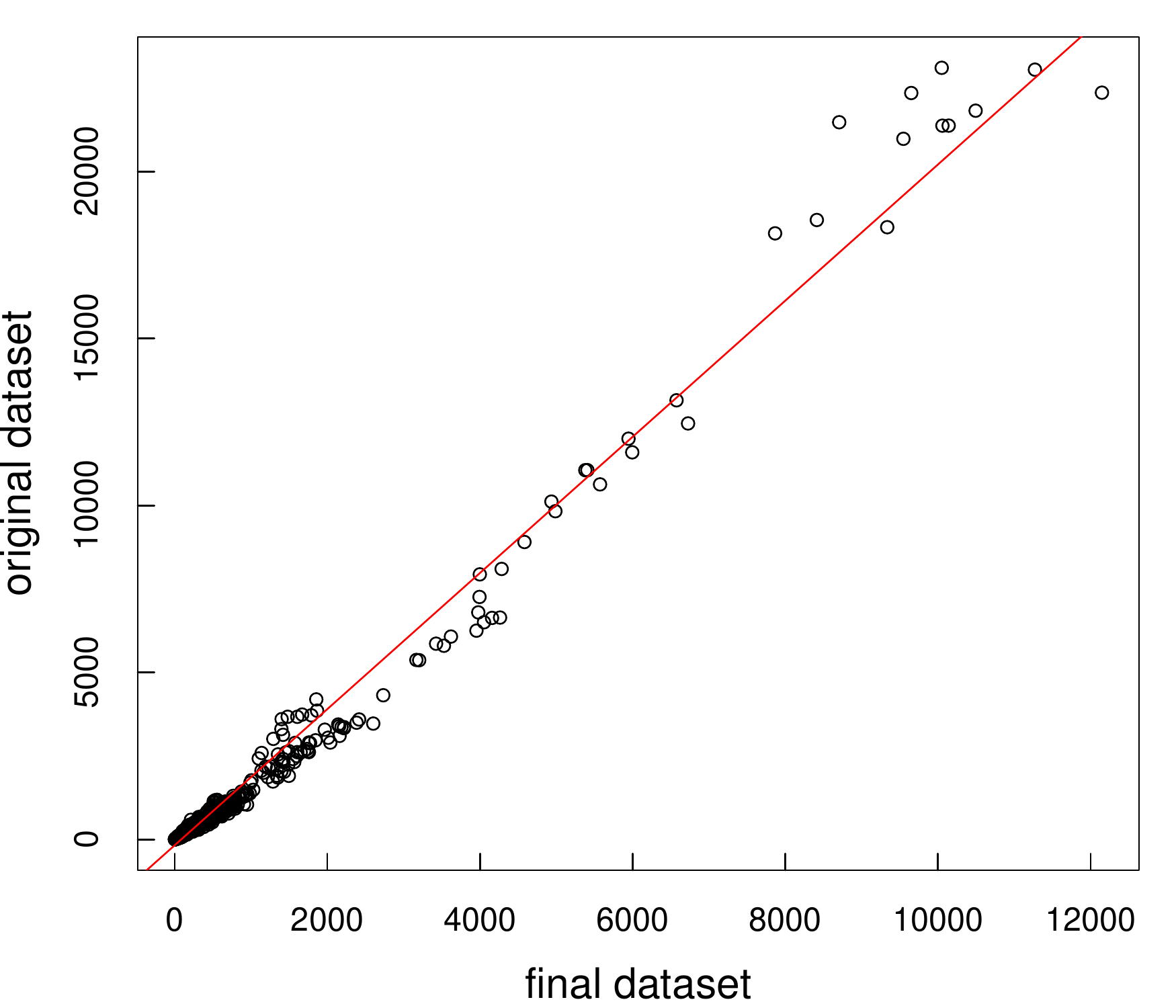} \label{fig:dwell_ads_transactions_city}}
 \hspace{0mm}
\subfigure[Delistings (y-o-y growth rates)] 
{\includegraphics[width=0.45\linewidth]{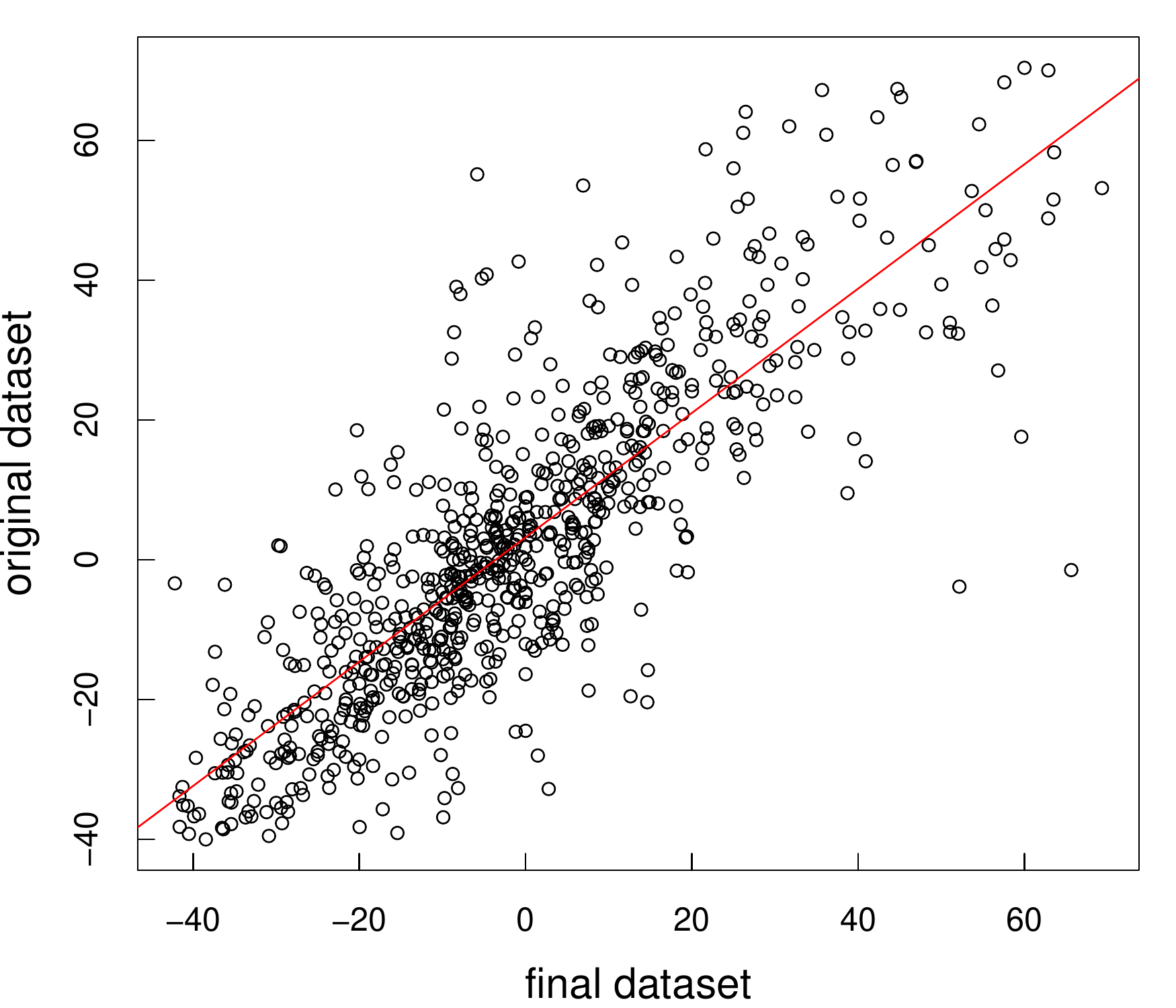}  \label{fig:dwell_ads_transactions_city_growth}}
\vspace{0mm}
\subfigure[Asking prices] 
{\includegraphics[width=0.45\linewidth]{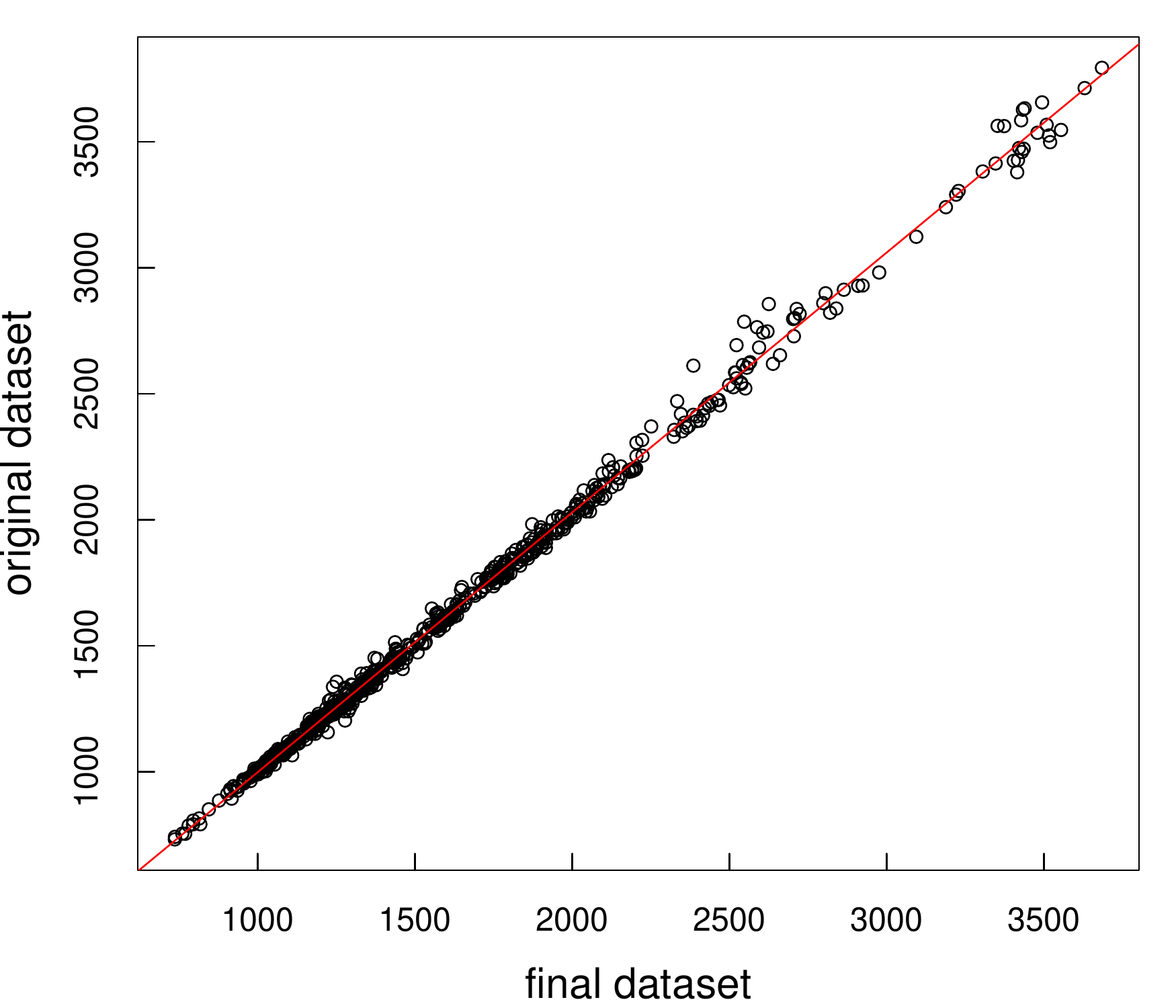} \label{fig:dwell_ads_prices_city}}
 \hspace{0mm}
\subfigure[Asking prices (y-o-y growth rates)] 
{\includegraphics[width=0.45\linewidth]{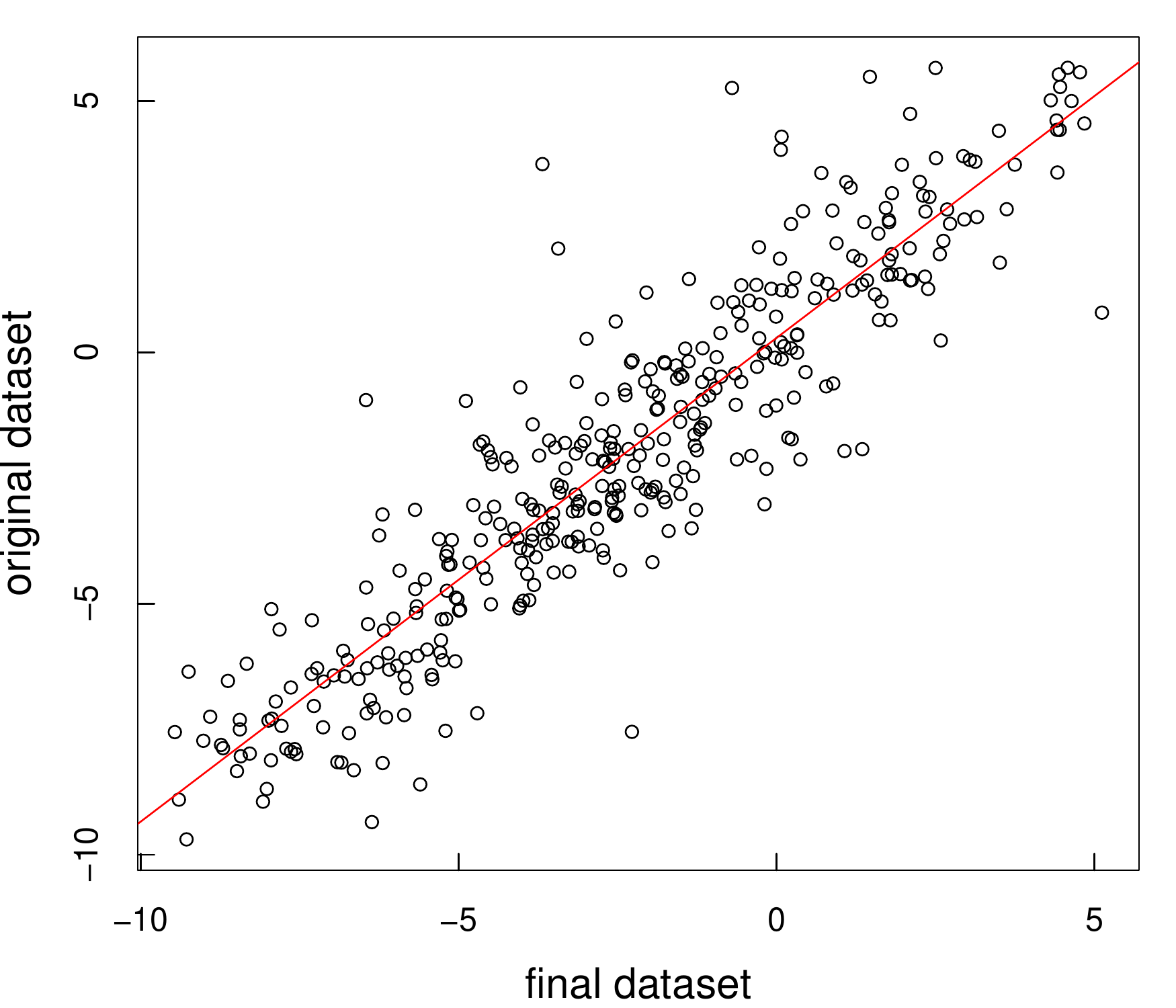}  \label{fig:dwell_ads_prices_city_growth}}
\caption{Original and final dataset (cities).}
\end{figure}
 
Figures \ref{fig:dwell_ads_transactions_city}-\ref{fig:dwell_ads_prices_city_growth} show the comparison for delistings and average listing prices when the level of aggregation is the city. We find an almost perfect correlation between the two datasets for both the number of delistings and average listing prices. When we calculate year-on-year rates of change, however, noise is stronger but correlation is still good. The correlation coefficients are 0.83 for asking prices and 0.68 for delistings. 

\begin{figure}[h]
\centering
\subfigure[Delistings] 
{\includegraphics[width=0.45\linewidth]{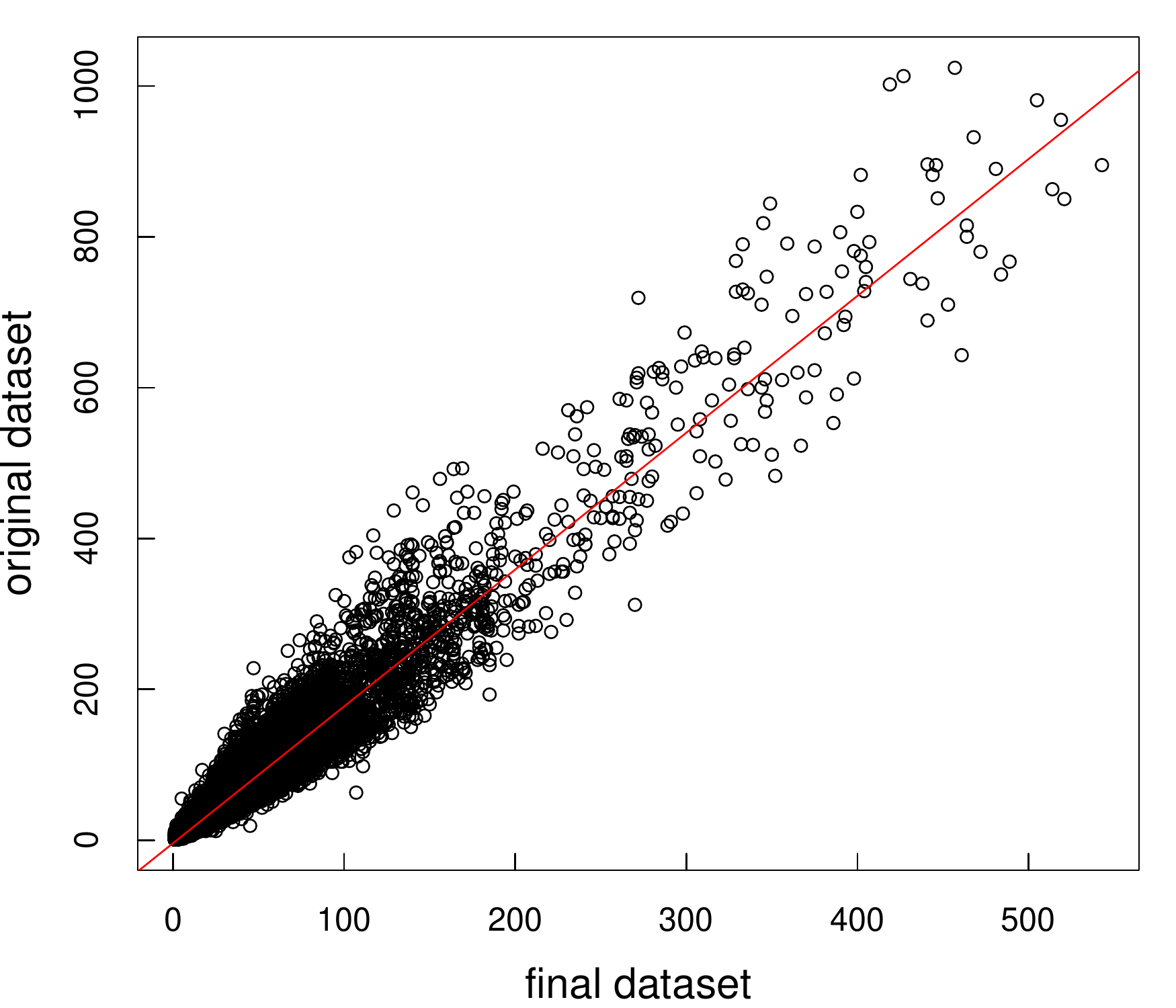} \label{fig:dwell_ads_transactions_zona}}
 \hspace{0mm}
\subfigure[Delistings (y-o-y growth rates)] 
{\includegraphics[width=0.45\linewidth]{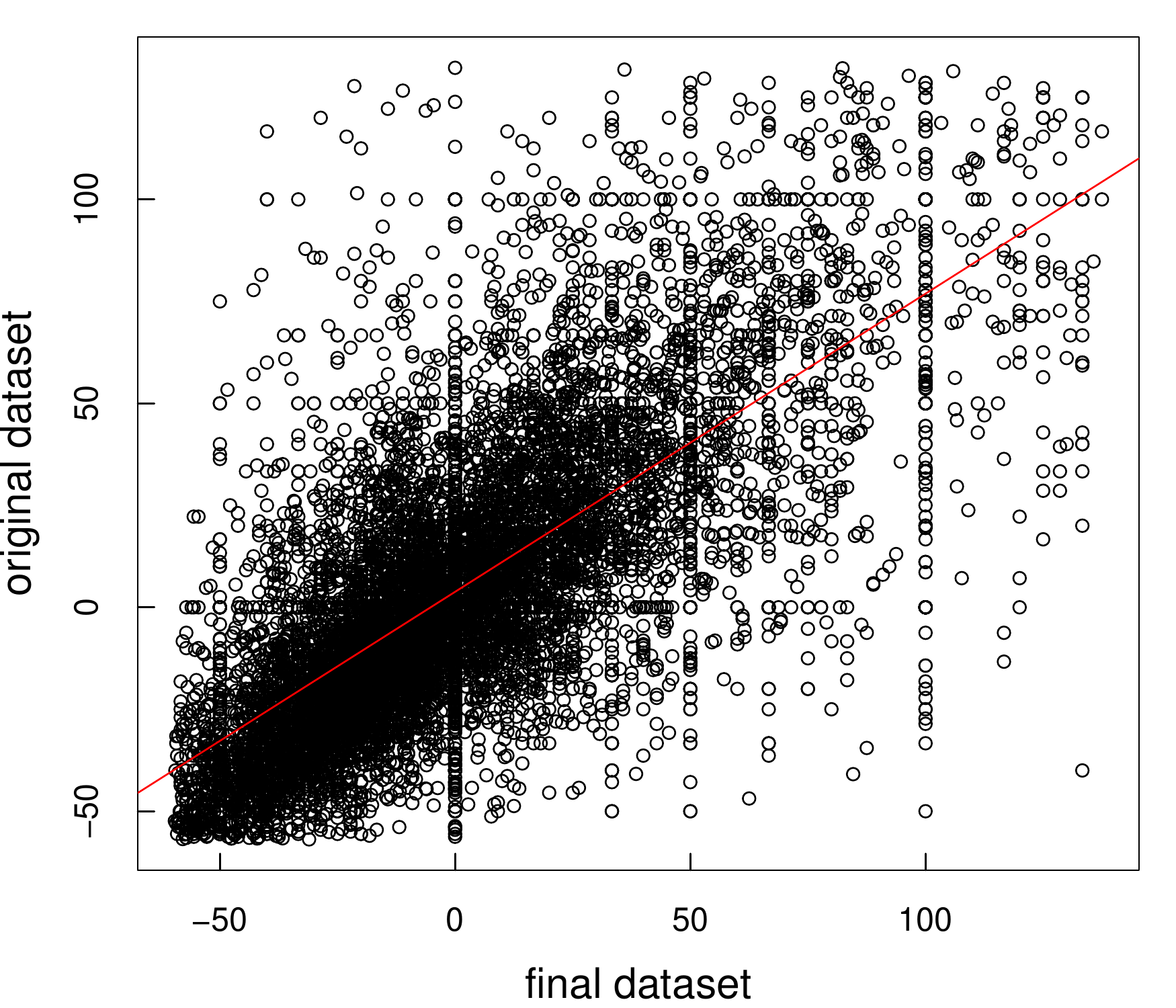}  \label{fig:dwell_ads_transactions_zona_growth}}
\vspace{0mm}
\subfigure[Asking prices] 
{\includegraphics[width=0.45\linewidth]{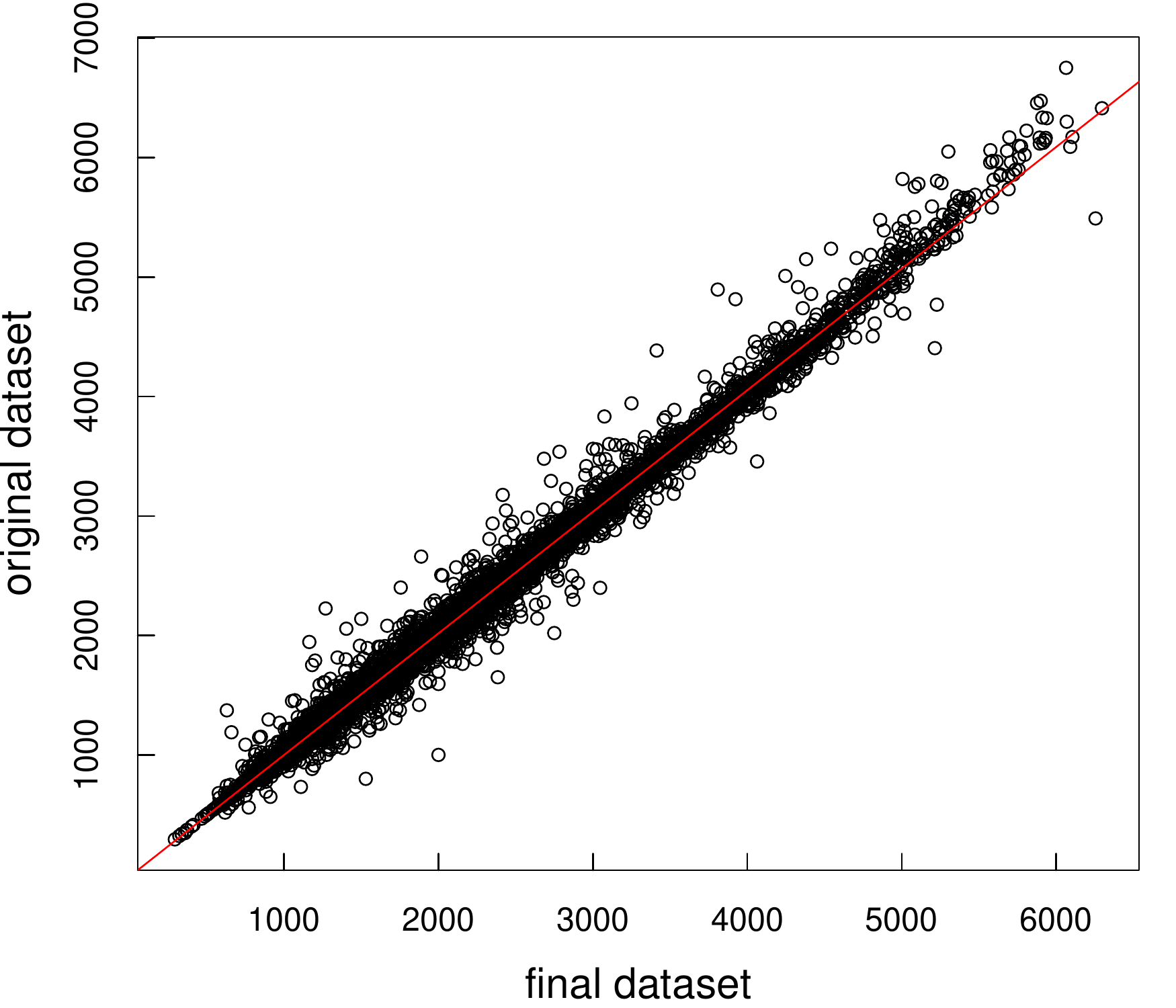} \label{fig:dwell_ads_prices_zona}}
 \hspace{0mm}
\subfigure[Asking prices (y-o-y growth rates)] 
{\includegraphics[width=0.45\linewidth]{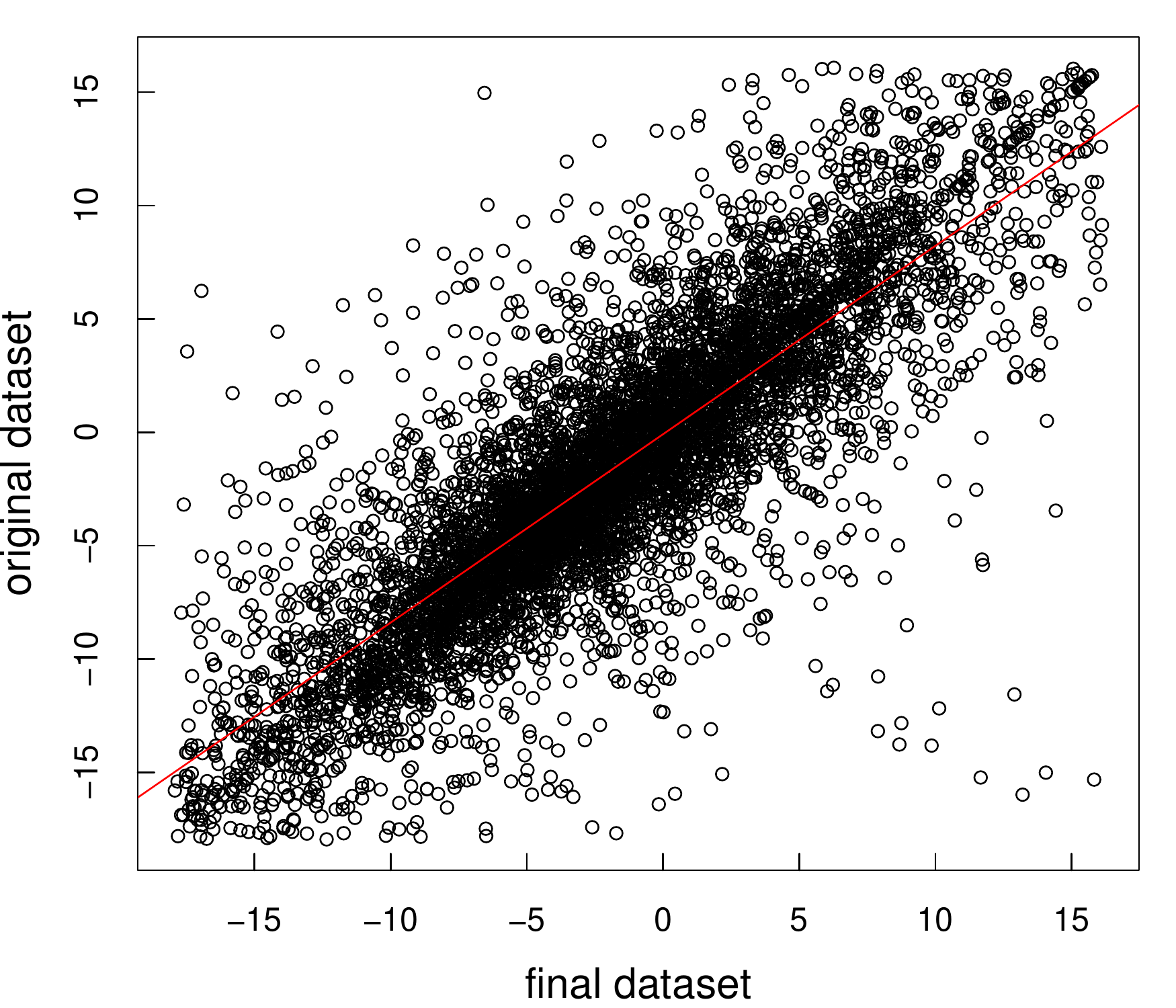}  \label{fig:dwell_ads_prices_zona_growth}}
\caption{Original and final dataset (local housing markets).}
\end{figure}

Figures \ref{fig:dwell_ads_transactions_zona}-\ref{fig:dwell_ads_prices_zona_growth} show that these correlations decrease when we consider the same variables at the local housing market level. In particular, when we consider year-on-year growth rates the correlation is relatively low for both variables (0.52 for transactions and 0.68 for prices). Figures \ref{fig:dwell_ads_transactions_zona_growth} and \ref{fig:dwell_ads_prices_zona_growth} show that in many cases the two datasets provide opposite indications about the evolution of the variable of interest compared to the previous year.

Therefore, the measurement error implied by keeping duplicate listings in the sample is sizable at granular level, in particular when we look at dynamics. However, depending on the application at hand, the original dataset of ads can be a reliable source of information. For example, an analyst may be interested only in the heterogeneity in prices across cities or the geographical level of aggregation is sufficiently broad. More generally, we suggest cautiousness in using online ads, especially in the estimation of the number of dwellings on the market and the composition of supply.

\subsection{Systematic bias}
\label{sec:duplicates_det}

Now, we test if it is more likely to observe duplicate ads for certain houses rather than others. In particular, we focus on two hypotheses. 

Our first hypothesis is that duplicate ads are more likely among those houses for which potential buyers show little interest, i.e. demand for these houses is relatively small. Intuitively, home sellers would choose to increase search intensity -- through open listing agreements with multiple agencies or more generally by posting numerous ads -- to compensate the scarcity of buyers potentially interested in their houses. 

Our second hypothesis is that the presence of duplicates is correlated with the listing price. It is reasonable that houses whose listing price is too high compared to similar nearby homes may have multiple associated ads because the seller increases his odds to find a buyer, for example by giving the mandate to sell to several agents.

To test for these two hypotheses we use weekly data to estimate the following linear probability model:\footnote{ We use a linear probability model instead of a logit model because of computational convenience.  }
\begin{align} \label{eq:dupl_1}
DUPL_{ijt} = \alpha_{jt} + \beta CLICKS_{ijt} + \gamma PRICE_{ijt} + \delta \mathbf{X}_i + \varepsilon_{ijt}  
\end{align}  
where $DUPL_{ijt}$ is an indicator variable, that is equal to one if more than one ad is associated to home $i$ in week $t$; the index $j$ refers to the local housing market in which the home is located. $CLICKS_{ijt}$ is average daily number of visits to the web pages (\textit{clicks}) related to dwelling $i$ during week $t$.\footnote{ When more than one ad refer to dwelling $i$, $CLICKS$ is computed in two steps. First, we compute the average daily number of clicks for each ad. Second, we compute the mean of the daily number of clicks across all ads. } We show in section \ref{sec:demand} that this measure proxies housing demand. Intuitively, the most searched houses are likely to be those for which the owner or the broker receive more calls or emails by potential buyers. $PRICE_{ijt}$ is the listing price per square meter of dwelling $i$ during week $t$. We control for spatial and temporal heterogeneity at the local housing market level through the set of dummies $\alpha_{jt}$. Finally, $\mathbf{X}_i$ is a vector with the following physical characteristics of dwelling $i$: floor area (square meters), type of property (apartment, detached dwelling), floor level, number of bathrooms, maintenance status, presence of a balcony or a terrace, garage and elevator.

\begin{table}[!ht] \centering 
 \footnotesize 
\begin{tabular}{@{\extracolsep{5pt}}lcccc} 
\\[-1.8ex]\hline 
\hline \\[-1.8ex] 
\\[-1.8ex] & Multiple ads & New duplicate & New duplicate & New duplicate \\ 
\\[-1.8ex] & (1) & (2) & (3) & (4)\\ 
\hline \\[-1.8ex] 
 Listing Price $t$ & 0.0198$^{***}$ &  &  &  \\ 
  & (0.0011) &  &  &  \\ 
  Clicks $t$ & $-$0.1221$^{***}$ &  &  & 0.4068$^{***}$ \\ 
  & (0.0010) &  &  & (0.0008) \\ 
  Clicks $t-1$ &  & $-$0.0015$^{***}$ & $-$0.0027$^{***}$ & $-$0.1816$^{***}$ \\ 
  &  & (0.0003) & (0.0003) & (0.0008) \\ 
  Clicks $t-2$ &  &  &  & $-$0.0429$^{***}$ \\ 
  &  &  &  & (0.0008) \\ 
  Clicks $t-3$ &  &  &  & $-$0.0211$^{***}$ \\ 
  &  &  &  & (0.0008) \\ 
  Clicks $t-4$ &  &  &  & $-$0.0154$^{***}$ \\ 
  &  &  &  & (0.0007) \\ 
  Clicks $t+1$ &  &  &  & $-$0.0156$^{***}$ \\ 
  &  &  &  & (0.0008) \\ 
  Clicks $t+2$ &  &  &  & $-$0.0559$^{***}$ \\ 
  &  &  &  & (0.0008) \\ 
  Clicks $t+3$ &  &  &  & $-$0.0299$^{***}$ \\ 
  &  &  &  & (0.0008) \\ 
  Clicks $t+4$ &  &  &  & $-$0.0310$^{***}$ \\ 
  &  &  &  & (0.0007) \\ 
  Listing Price $t-1$ &  & 0.0024$^{***}$ &  & 0.0052$^{***}$ \\ 
  &  & (0.0004) &  & (0.0005) \\ 
  Listing Price $t-4$ &  &  & 0.0013$^{***}$ &  \\ 
  &  &  & (0.0002) &  \\
   Fixed effects & OMI microzone +  & OMI microzone +  & OMI microzone + & OMI microzone + \\
   & week & week & week & week \\  
 Observations & 16,042,720 & 15,450,398 & 14,374,903 & 11,330,074 \\ 
Adjusted R$^{2}$ & 0.0036 & 0.0004 & 0.0004 & 0.0231 \\ 
\hline \\[-1.8ex] 
\multicolumn{5}{l}{\textit{Notes:} Coefficients and standard errors reported in the Table have been multiplied by 100.}  
\end{tabular} 
\caption{Determinants of duplicates} 
  \label{table:dupl_determinants} 
\end{table}

Column (1) in Table \ref{table:dupl_determinants} reports the results of our estimation. The coefficients associated with $CLICKS$ and $PRICE$ are statistically significant, and their sign confirms our initial hypotheses. The presence of multiple ads is associated with low interest by potential buyers, and a relatively higher listing price. Although we cannot claim any causal relation based on model \eqref{eq:dupl_1}, the evidence is consistent with the hypothesis that the home seller increases his effort to find a buyer to compensate for a high asking price or unattractive characteristics of the home.   

To identify a causal effect of demand and listing prices on the propensity to post multiple ads, we create a new indicator variable called $NEWDUPL$. This variable is equal to one if the number of ads associated with a home already on the market increases during week $t$. Then, we estimate the following linear probability model:
\begin{align} \label{eq:dupl_2}
NEWDUPL_{ijt} = \alpha_{jt} + \beta CLICKS_{ijt-1} + \gamma PRICE_{ijt-1} + \delta \mathbf{X}_{i}+\zeta z_{it} + \varepsilon_{ijt}  
\end{align}  
Compared to \eqref{eq:dupl_1}, we take as regressors the one-week lag for both demand and listing price. This model allows us to test if the propensity of the home seller or the broker to increase advertising during week $t$ by posting a new ad is affected by asking price and buyers' demand during the previous week. We also control for the number of days dwelling $i$ has been listed up to week $t$ ($z_{it}$). 

Column (2) in Table \ref{table:dupl_determinants} shows that our previous results are qualitatively confirmed. The propensity to post a new ad for a previously listed home decreases when online interest for that home goes up; this propensity is also increasing in the listing price. Notice that these coefficients are statistically significant although we include many controls and the phenomenon we are considering is not very frequent at weekly frequency. In particular, the unconditional probability that during week $t$ a new ad is posted for a previously listed home is 0.9 percent.  

In this regression, clicks can be considered as exogenous because potential buyers cannot know the sellers' strategies a week in advance. Moreover, since we control for the listing price and dwellings characteristics, we deduce that the lower online attention is determined not only by an excessively high price asked by the seller but by a genuine mismatch with potential buyers' preferences. Unfortunately, we cannot resort to this argument to claim that the lagged value of the listing price is exogenous. Indeed, home sellers/brokers set both the listing price and the advertising strategy, and when changing the listing price they can have already decided to post a new ad. However, we show in column (3) that replacing the one-week lagged listing price with the four-weeks lag, we still find a positive and significant effect on the propensity to post a new ad. 

Finally, after having shown that the listings that receive little online attention are those with the highest probability of having multiple ads, we want to evaluate the effectiveness of this advertising strategy. To do that, we estimate the following extension of model \eqref{eq:dupl_2}:
\begin{align} \label{eq:dupl_3}
NEWDUPL_{ijt} = \alpha_{jt} + \sum_{i=-4}^4 \beta_i CLICKS_{ijt-i} + \gamma PRICE_{ijt-1} + \delta \mathbf{X}_{i}+\zeta z_{it} + \varepsilon_{ijt}  
\end{align}
where we add as regressors the contemporaneous value of the variable $CLICKS$ and all its lags and leads up to four weeks. The results are reported in column (4) in Table \ref{table:dupl_determinants}. We find that during the four weeks before the seller posts a new ad, his home gets a relatively poor online interest ($\beta_i <0$ for $i=-1,-2,-3,-4$). Clicks are low especially in the previous week ($\beta_{-1}$). Then, following the publication of the new ad, a spike in clicks occurs ($\beta_{0}>0$). That happens because potential buyers may believe that this is a new listing. Home buyers may not easily recognize that the new ad refers to a previously listed house, and this is especially true when a new broker posts the ad. However, this positive effect is only temporary: In the following four weeks, online attention returns to relatively low levels again ($\beta_i <0$ for $i=1,2,3,4$). 

The main conclusion is that using the original dataset of ads implies an oversampling of houses relative expensive -- given their location and characteristics -- and less attractive. An implication is that by using ads we incur into the overestimation of the average level of the listing prices. Moreover, houses with multiple ads show further systematic differences compared to other dwellings. We estimate the OLS regression of time on market over a dummy taking value one if a home had multiple ads, and we find that those with many ads stay longer on the market (see Table \ref{table:dupl_tom}). We also estimated a linear probability model where the dependent variable is an indicator variable taking value one if the home seller revised downward the initial asking price, and zero otherwise. As expected, it is more plausible to observe a price change for houses with multiple ads (Table \ref{table:dupl_tom}). These results are consistent with previous evidence: by using the ads, we underestimate the time on the market, and dwellings with multiple ads are overpriced (therefore more subject to price reductions).

\begin{table}[!htbp] \centering 
 \footnotesize 
\begin{tabular}{@{\extracolsep{5pt}}lcc} 
\\[-1.8ex]\hline 
\hline \\[-1.8ex] 
\\[-1.8ex] & Time on market & Price change \\ 
\\[-1.8ex] & (1) & (2)\\ 
\hline \\[-1.8ex] 
 Multiple ads & 125.30580$^{***}$ & 0.17805$^{***}$ \\ 
  & (0.74404) & (0.00150) \\ 
  Fixed effects & OMI microzone & OMI microzone \\
  Temporal effects & Quarter & Quarter \\
 Observations & 512,246 & 512,246 \\ 
Adjusted R$^{2}$ & 0.06827 & 0.09316 \\ 
\hline \\[-1.8ex] 
\end{tabular} 
\caption{Duplicates, time on market and price changes} 
  \label{table:dupl_tom} 
\end{table}

\FloatBarrier

\section{Measuring demand, supply and prices}
\label{sec:measure_stats}

So far we analyzed the potential measurement problems associated with the use of online listing data for the study of housing markets, and have outlined a procedure to overcome the main critical issues. In the second part of the paper we analyze the potential of these data and illustrate their complementarity with traditional statistical sources.

\subsection{Demand}
\label{sec:demand}

Online listing data allow to extract information about housing demand. Indeed, online activity leaves digital traces of human behavior. When individuals visit the webpage of an ad, they convey information about the characteristics, the location and the price interval of the home they are searching for. By aggregating all this information, we can understand which area households are searching more intensely and what they are looking for. We can observe housing demand.

Here, we know how many times website users visited the web page of an ad during each week (clicks). This information is complementary with respect to web searches, namely queries where the user specifies the location, characteristics and price range of the home she is looking for.\footnote{ Web searches are for example used in \citealt{piazzesi_schneider_stroebel_2015}. }  

In principle, web searches and clicks do not convey the same information. People may search for houses with a bundle of characteristics that cannot be found in the market. In this case, we cannot observe clicks that map to those preferences, and so, we do not observe the actual preferences of potential buyers. At the same time, we think that for a large class of applications there is no loss of generality in using clicks instead of web searches. For example, if an individual is really interested in buying a house, it is plausible that it would somewhat adapt its preferences to the composition of the supply. 

Clicks are also more convenient than web searches for analysis. First, home listing websites usually allow ``map search'', letting users specify a polygon on the map where they look for a home. Extracting and aggregating this type of information about buyers preferences is definitely hard \citep{piazzesi_schneider_stroebel_2015,rae_2015}, while this problem does not arise when considering visits to web pages. Besides, clicks are available for each listing, and they can be used to proxy the interest of potential buyers for each home.  

To show that online interest is a proxy of housing demand, we test whether online interest for a dwelling is correlated with the time it has been on the market and the occurrence of price revisions. If the webpage of a listed dwelling gets many clicks, it is plausible that many households are searching for that type of houses. Therefore, our first hypothesis is that high online interest is associated with shorter time on market. Moreover, it is plausible that the price interval is a key searching criterion which is set by all potential buyers. If many households search in a given price interval for a dwelling with a particular bundle of characteristics, \textit{ceteris paribus} it is less plausible to observe downward revisions of the asking price for these dwellings. This is our second hypothesis.       
  
We consider dwellings that have been listed since the beginning of 2016, and we keep only those with no duplicate ads, to avoid to deal with the bias identified in Section \ref{sec:duplicates_det}. We introduce a new variable $ONLINT$ to quantify the relative interest in a particular dwelling over its listing history with respect to all other dwellings in the same local housing market.\footnote{ We cannot use the variable $CLICKS$ because houses are listed at different times and for different periods. } $ONLINT$ is computed as the total number of clicks on a house, divided by the average number of clicks in the corresponding local housing market during the same period in which the home has been on the market. We run the following OLS regression:
\begin{equation*}
 \log(TOM_{ijt}) =  \alpha_j + \zeta_t + \beta \log (ONLINT_{ijt})  +  \delta \mathbf{X}_{ijt} +  \varepsilon_{ijt},
\end{equation*}
where $TOM$ is the time on market -- measured as the number of days between the delisting and the first listing -- of dwelling $i$ in the local housing market $j$ delisted in quarter $t$. The vector $\mathbf{X}$ includes the physical characteristics of the dwelling and the relative asking price per square meter, defined as the ratio of the price per square meter of the dwelling to the average price per square meter in the local housing market  over all the period when the dwelling was listed. We add local housing market and quarter dummies, to capture spatial and temporal effects. In column (1) of Table \ref{tab:demand_regressions}, we report the results of this regression. The coefficient associated to $ONLINT$ is both statistically and economically significant, and its sign confirms our hypothesis. According to our estimate, a 1\% higher number of clicks is on average associated with a 0.6\% shorter time on market. 

We now test whether online interest predicts the occurrence of price revisions. Since price revisions likely trigger a change of online interest, we slightly change the definition of $ONLINT$ to avoid incurring in this issue. We compute the variable $ONLINT2$ as the ratio of clicks in the first 14 days in which the dwelling was listed to the average of the local housing market in the same period. We introduce a binary variable $PRICEREF$, that takes value one if the asking price of the dwelling was revised downward, and zero if it was not revised or if it was revised upward.\footnote{ We consider only the case of downward price revisions for two reasons. First, the number of upward revisions is relatively small. Second, it is not always clear from the ads if a price increase is due to an improvement of the conditions offered by the seller or to some unobserved change in the quality of the dwelling, such as the inclusion of the garage in the sale. } Then, we run the following logistic regression:
\begin{equation*}
 \log \left(\frac{p_{ijt}}{1-p_{ijt}} \right) =  \alpha_j + \zeta_t + \beta ONLINT2_{ijt}  + \delta \mathbf{X}_{ijt} + \varepsilon_{ijt},
\end{equation*}
where $p \equiv Prob(PRICEREF=1)$ and, as in the previous regression, we control for the relative asking price per square meter, the physical characteristics of the dwelling and local housing market and quarter dummies. We estimate that a 1\% increase in the relative number of clicks is associated with a 12\% reduction in the odds of a downward price revision (Table \ref{tab:demand_regressions}, column 2).\footnote{In \cite{pangallo_loberto_2018} we show that the relation between prices and online interest works also the other way around. We find that a 1\% higher price is associated to a 0.66\% lower number of clicks. We also show that this elasticity has a causal interpretation.}

\begin{table}[!ht] \centering 
\footnotesize 
\begin{tabular}{lcccc} 
\hline 
\hline
\\ 
 & \multicolumn{4}{c}{\textit{Dependent variable:}} \\ 
\cline{2-5} 
\\ 
& $TOM$ & $PRICEREV$ & $AVPRICE$  & $LIQUIDITY$ \\ 
 & (1) & (2)& (3) & (4) \\ 
\hline \\ 
  $ONLINT$ & -0.599$^{***}$ &   &  & \\ 
  & (0.002) &     &  & \\
  $ONLINT2$ &   &   -0.120$^{***}$ &  & \\ 
  &  &   (0.010) &  & \\
  $DEMAND_{t-1}$ & & & 0.033$^{**}$ &  0.060$^{***}$ \\ 
 & & & (0.015) &  (0.018)  \\ 
  $DEMAND_{t-2}$ & & & 0.048$^{***}$ & 0.138$^{***}$    \\ 
  & & & (0.015) &  (0.018)   \\ 
  $AVPRICE_{t-1}$ & & & 0.528$^{***}$ &   \\ 
  & & & (0.009) &   \\ 
  $LIQUIDITY_{t-1}$ & & &  & 0.161$^{***}$  \\ 
  & & &   & (0.011)  \\ 
  Fixed effects & OMI microzone & OMI microzone & City & City \\
  Temporal dummies & Quarter & Quarter & Quarter & Quarter \\  
 \hline \\ 
Observations & 201,208 &  200,000 & 8,851 & 8,851 \\
$R^2$ & 0.36  &  -- & 0.54 & 0.35 \\ 
\hline 
\hline \\
\end{tabular} 
  \caption{Online interest} 
  \label{tab:demand_regressions} 
\end{table}

Next, we test if online interest predicts aggregate housing market dynamics. We build an indicator of housing demand in each local housing market, and we expect that aggregate online interest is correlated with market liquidity and house prices. Our hypothesis is that stronger demand should lead to a more liquid market and to an increase in house prices, as suggested in the recent literature \citep{carrillo_etal_2015, wu_brynjolfsson_2015,vandijk_francke_2017}.

We construct the variable $DEMAND$, defined as the average number of clicks per ad in a local housing market. Then, we test whether aggregate online attention is a leading indicator of average asking prices ($AVPRICE$) and liquidity ($LIQUIDITY$), mostly following the approach of \cite{vandijk_francke_2017}.\footnote{ \cite{vandijk_francke_2017} analyze a dataset of online housing ads in the Netherlands and show that the average number of clicks Granger-causes liquidity and prices. We confirm their findings, and extend their analysis by considering smaller geographical aggregates -- the spatial unit in their analysis is a municipality, here we consider local housing markets. } We aggregate our data at quarterly frequency and consider two time lags for our measure of online interest. As control variables, we consider $DEMAND$ with one and two lags, city and time dummies. In addition, we use socio-economic characteristics of the local housing markets from the 2011 census ($\mathbf{X}$). These variables are the fraction of people with a university degree, the unemployment rate, the fraction of owner-occupied houses and the percentage of foreign population. Market liquidity ($LIQUIDITY$) is measured as the ratio between the number of dwellings leaving the market (delistings) and those on the market in the same period (total number of listings). We run the following OLS regression:
\begin{align*}
 \log(Y_{ijt}) = &\alpha_j + \zeta_t + \beta_1 \log (DEMAND_{ijt-1})+ \beta_2 \log (DEMAND_{ijt-2})  +\\ &\beta_3 \log (Y_{ijt-1})   + \delta \mathbf{X}_i + \varepsilon_{ijt},
\end{align*}
where $Y_{ijt}$ is $AVPRICE$ or $LIQUIDITY$ for the local market $i$, located in city $j$ in quarter $t$. The results reported in columns (3) and (4) of Table \ref{tab:demand_regressions}  confirm that online interest is a good leading indicator of prices and liquidity.

Therefore, clicks to webpages of the ads are an effective tool to measure housing demand in real time and to understand buyers' preferences. Moreover, differently from buyers' web searches, clicks are easy to handle, and they allow to build a measure of demand for a specific house, not only for a neighborhood or a typology of dwellings.

\FloatBarrier

\subsection{Supply}
\label{sec:supply}

Housing supply is usually defined as the total number of dwellings, independently on whether they are on sale or not \citep{glaeser_gyourko_2018}. As a consequence, an increase in housing supply is represented by new constructions, and housing supply is downward rigid because of the durable nature of dwellings. 

In the short or medium-run this definition is not necessarily the most suitable. Indeed, the number of houses potentially available for sale changes over time, at least for two reasons. First, a significant fraction of home listings are currently inhabited by their owners that are contemporaneously searching as buyers. Since the decision to move can depend on macroeconomic developments and on the housing-market conditions, the actual supply of dwellings changes across the cycle and is not a fixed fraction of the stock of dwellings \citep{anenberg_bayer_2013,ngai_sheedy_2018}. Second, the housing market consists of two main segments: the market for property and the market for renting. Owners of vacant houses always have the option either to search for a buyer or for a tenant  \citep{krainer_2001, head_etal_2014, liberati_loberto_2019}.

\begin{table}[!ht] \centering 
\footnotesize 
\begin{tabular}{lcccc} 
\hline 
\hline
\\ 
 & \multicolumn{4}{c}{\textit{Dependent variable:}} \\ 
\cline{2-5} 
\\ 
& $FLOORAREA$ & $BATH$ & $GARDEN$  & $TERRACE$ \\ 
 & (1) & (2)& (3) & (4) \\ 
\hline \\ 
  $HEDON$ &  0.208$^{***}$ & 20.212$^{***}$  &  12.996$^{***}$   & 13.363$^{***}$  \\ 
  & (0.039) &  (3.033) & (2.321)  & (3.072) \\  
  Fixed effects & City & City & City & City \\
  Temporal dummies & Quarter & Quarter & Quarter & Quarter \\ 
 \hline \\ 
Observations & 1,064 &  1,064 & 1,064 & 1,064 \\
$R^2$ & 0.93  &  0.96 & 0.97  & 0.98 \\ 
\hline 
\hline \\
\end{tabular} 
  \caption{Quality of listed dwellings and house prices } 
  \label{tab:supply_regressions} 
\end{table}

As a result, the number and type of houses that are offered for sale on the market may not correlate with the total number of houses, and for some applications it is more reasonable to look at listings as the housing supply.\footnote{It is fair to say that this distinction is the same that arises in labour market statistics, in which only people that are already working or searching actively for a job are considered inside the labour supply.} For example, \cite{ngai_sheedy_2018} show that home sales variation is mostly driven by listings instead of change in matching efficiency in the housing market. Moreover, also the composition of housing supply changes over the housing market cycle. 

To show that the average quality of the houses offered for sale changes with the timing of the real estate cycle, we consider four variables that measure average quality of listings in each local housing market at quarterly frequency. We define $FLOORAREA$ the logarithm of the average floor area of listings (measured in square meters); $BATH$ is the share of listings with at least two bathrooms; $GARDEN$ is the share of listings having a private garden; $TERRACE$ is the share of listings having a terrace. To meaningfully compute  these variables, we consider only local housing market with at least 50 listings during each quarter. As a measure of the timing of the housing market cycle in each city we use the logarithm of hedonic house asking price index ($HEDON$). We consider this variable because changes in hedonic prices are by construction uncorrelated with changes in home average quality. Finally, to show that the composition of home supply changes with house prices, and that it does not depend on the characteristics of newly built houses, we consider only existing dwellings.

We estimate the following model
\begin{equation}
Y_{ijt} = \alpha_j + \zeta_t + \beta HEDON_{jt}+\varepsilon_{ijt}
\end{equation}
where the dependent variable $Y$ is one among $FLOORAREA$, $BATH$, $GARDEN$ and $TERRACE$. We add city dummies -- to control for the heterogeneity of the housing stock in the different cities -- and quarter dummies. We report the results in Table \ref{tab:supply_regressions}. Housing supply in cities with house price increases is characterized by a larger average floor area of listed houses and greater shares of listings with at least two bathrooms, a private garden and a terrace. Therefore, house price increases seem to be associated with a better quality of housing supply.

\FloatBarrier

\subsection{Prices}
\label{sec:prices}
As first documented by \cite{anenberg_laufer_2017},  listing data provide several advantages in measuring house prices. First, listing prices are observed in real time, while sale prices are usually available with a significant lag. Second, buyers and sellers agree about the sale price at the time of delisting, while the settlement date -- that is the relevant date for computing a house price index --  is usually several weeks later. Due to this lag, listing prices can predict a standard house price index over a short term horizon.

Here, we use our dataset to first discuss a potential problem regarding the use of listing data to measure house prices, and then to highlight a possible advantage.

The potential problem is that the growth rate of asking prices is a good approximation for the rate of change of sale prices only if the average discount with respect to asking prices obtained by buyers in the bargaining process were stable.\footnote{ We can express the relation between asking prices, $P^a_t$, and sale prices, $P_t$, as $P^a_t = \left( 1-d_t \right)P_t$, where $d_t$ is the average discount. The dynamic relation between asking and sale prices is therefore given by $ P^a_t - P^a_{t-1} = P_t - P_{t-1} - \left( P_t d_t - P_{t-1}d_{t-1} \right)$. } Unfortunately, the outside option of both buyers and sellers in the bargaining process is affected by the general market conditions. Therefore, the average discount on asking prices changes over time.\footnote{ According to the Italian Housing Market survey the average spread between asking and effective prices increases during market downturns and decreases during market recoveries (see figure \ref{fig:tom_discount} in Appendix \ref{sec:addtabfig}). \cite{genesove_mayer_2001} show similar evidence for the United States, and they argue that this behavior is motivated by the fact that sellers suffer loss aversion. } As a consequence, a decrease in the discount implies that sale prices decrease less or increase more than asking prices. Potentially, in some periods sale prices may increase while asking prices decline. 

\begin{figure}[ht]
\centering
\subfigure[All cities] 
{\includegraphics[width=0.48\linewidth]{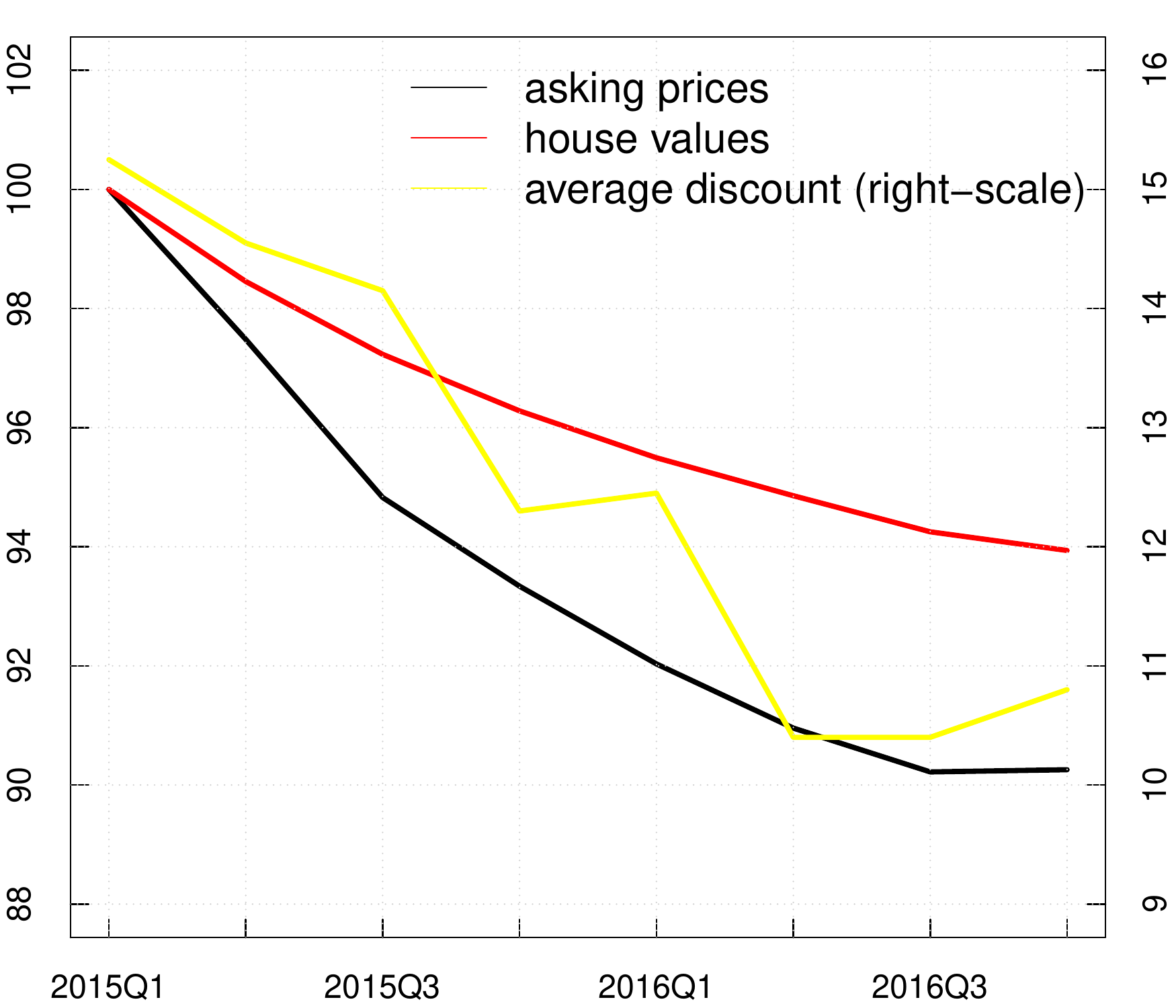} \label{fig:valid_pricesCF}}
 \hspace{0mm}
\subfigure[Rome and Milan] 
{\includegraphics[width=0.48\linewidth]{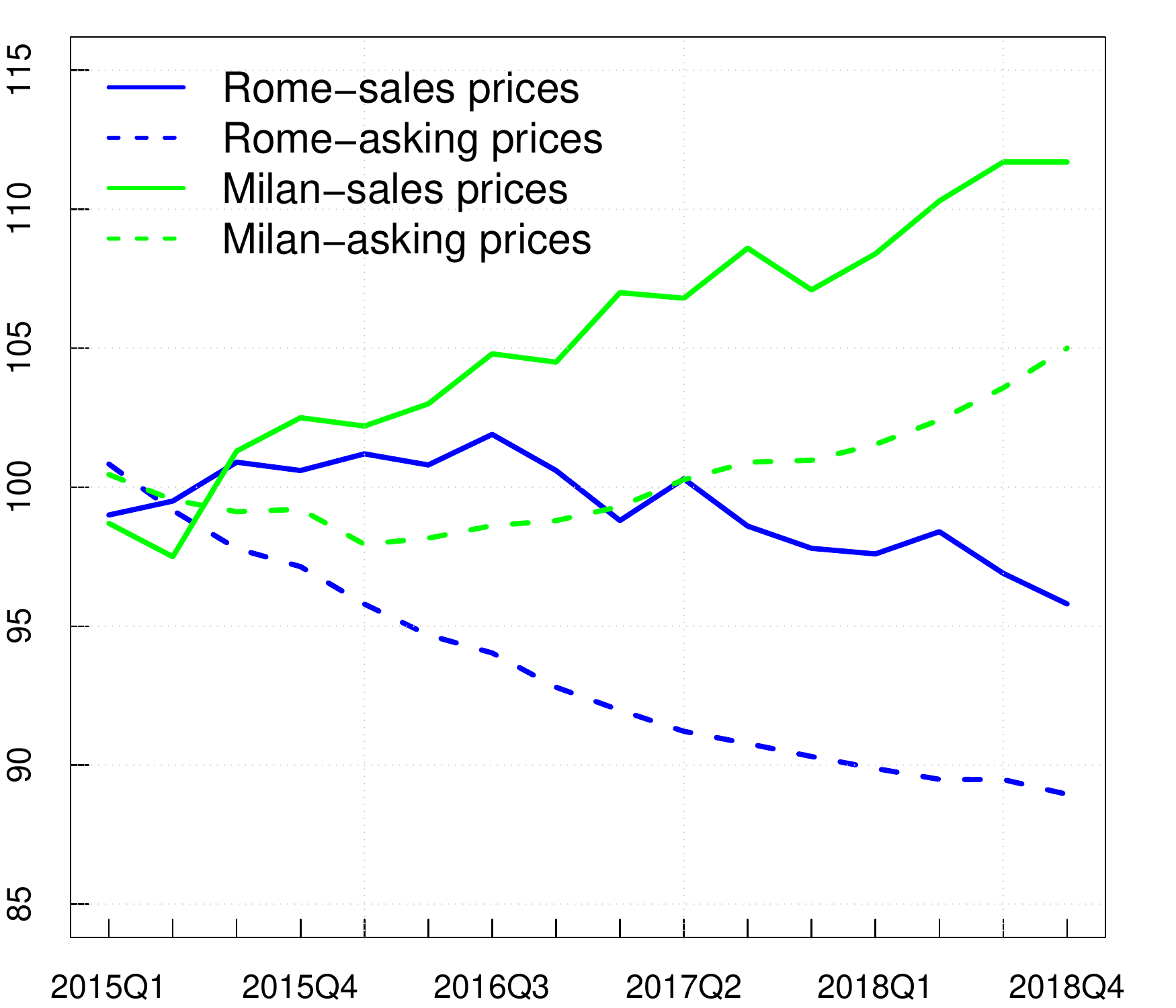}  \label{fig:ind_prices_miro}}
\caption{Prices (index 2015S1=100) and average discount (percentage points). }
\end{figure}

We illustrate this issue in Figure \ref{fig:valid_pricesCF}. Between the first semester of 2015 and the second semester of 2018, average home values (red line) declined by about 6.0\%, while asking prices (black line) diminished by 9.8\%. That is consistent with the observation that in Italy the average discount on asking prices (yellow line) -- estimated by the Italian Housing Market Survey -- decreased cumulatively by 4.4 percentage points over the same period. Thus, using asking prices to predict sale prices may require a sound estimate of the average discount.\footnote{ \cite{anenberg_laufer_2017} discuss this issue at the end of their paper. They show that including variables that are correlated with the discount -- such as time on market -- improves the forecasting performance of listing prices. }  

The possible advantage of using listing prices in place of sale prices starts from the consideration that we do not always need to predict the short-term dynamics of sale prices. More often, the goal is real-time identification of turning points in the house price cycle, i.e. if house prices are going to start a lasting recovery after a decline (or a severe downturn after a prolonged increase). Asking prices may be more reliable than sale prices for this real-time forecasting problem. An example is provided by the dynamics of (hedonic) asking and sale prices in the two main Italian cities, namely Rome and Milan (Figure \ref{fig:ind_prices_miro}). In 2016 sale prices increased, after a prolonged decline, both in Rome and Milan. Then, in Milan house prices continued to increase, while in Rome prices started to fall in 2017, as they were before 2016. Looking to asking prices it would have been possible to predict in real time that the 2016 sale price increase in Rome was temporary (asking prices were falling by almost 5\% year-on-year), while in Milan the resilience of the recovery was confirmed by the increase in asking prices. 

Despite this evidence is almost anecdotal given the small sample size,\footnote{We are unable to consider more cities as the sale price index provided by ISTAT is only available at the city level for Milan, Rome and Turin, and in the case of Turin it is extremely noisy.} we think that the higher reliability of asking prices for real-time identification of turning points in the house price cycle is a more general phenomenon, for the following reasons. First, when considering hedonic price indexes, listing data provide rich information about the physical characteristics of the housing units, while transaction data often contain little information.  Second, the larger volume of listings compared to sales allows to take into account a larger portion of the housing market, including homes that may be sold in the near future, thus smoothing short-term trends. Third, and finally, the short-run volatility of sale prices can be driven by the average discount. Indeed, Figure \ref{fig:ind_prices_miro} shows that, when the average discount stabilized, asking and sales started again having similar dynamics.

\FloatBarrier

\section{Conclusion}
\label{sec:conclusions}

We study a large dataset of housing ads, published on the main online portal for real-estate services in Italy. We provide a comprehensive analysis of the strengths and weaknesses of these data for the study of housing markets. The main issue is the existence of a substantial share of duplicate ads, which can lead to a misrepresentation of the volume and composition of housing supply. As examples of the potential of these data, we focus on three new applications on housing demand, supply and prices.

Although our analysis is of course specific to the dataset we use, we think that our insights could be employed more generally as economists increasingly rely on online listings websites. For example, the issue of duplicates is likely to affect all listings websites that have no incentive to control proliferation of duplicates, e.g. because they profit from the number of ads rather than from data quality. In the case of home listings websites, this problem is exacerbated by open mandate agreements: in all countries where these agreements are possible, duplicate ads could arise from different agencies, and we find it unlikely that website administrators could correct for this bias. 

Our paper provides a method to identify these duplicates and extract useful economic information on behavior of sellers in housing markets from this ``data cleaning'' procedure.

\newpage

\nocite{*}

\bibliographystyle{aea}
\bibliography{references}
\newpage
\appendix

\section{Data}
\label{sec:app_description_data}

\noindent \textbf{Listings}. The source data which we obtained from Immobiliare.it are contained in weekly files. Starting from these snapshots, we construct six datasets. The main dataset is the one with unique ads. Three datasets track the weekly change of asking prices, visits and uses of the form to contact the agency that is shown on each ad (we do not use information on contacts in this paper; in \cite{pangallo_loberto_2018} we show that it provides equivalent information to the number of visits). The last two datasets contain information about real estate agents and the the list of hash codes of the pictures associated to each ad.  The information available for each ad is reported in Table \ref{tab:content}.

\begin{table}[htbp]
	\centering
		\begin{tabular}{|lp{10cm}|}
			\hline
			 \textbf{Type of data} & \textbf{Variables}  \\		
			 \hline
			 Numerical & Price, floor area, \textit{rooms}, \textit{bathrooms}   \\
			\hline	
			Categorical & Property type, furniture, kitchen type, heating type, \textit{maintenance status}, \textit{balcony}, \textit{terrace}, \textit{floor}, air conditioning, energy class,  \textit{basement}, \textit{utility room}  \\
			\hline	
			Related to the building & \textit{Elevator}, \textit{type of garden}, \textit{garage}, \textit{porter}, building category  \\
			\hline	
			Contractual & \textit{Foreclosure auction}, contract type  \\
			\hline	
			Related to the seller & Publisher type (private citizen or real estate agency), agency name and address  \\
			\hline
			 Visual & Hash codes of the pictures, pictures count  \\
			\hline
			Geographical & Longitude, latitude, address \\
			\hline
			Related to the ad & Visits, contacts  \\
			\hline		
			 Temporal & Ad posted, ad removed, ad modified \\	
			\hline
			 Textual & Description \\
			\hline						
		\end{tabular}
		\caption{Information contained in the database provided by Immobiliare.it. For a complete description of the meaning of the variables, see \cite{loberto_etal_18}. Italics indicates that if variables are missing, we perform semantic analysis on the textual description of the ads to recover missing information.}
	\label{tab:content}
\end{table}

\FloatBarrier

\noindent \textbf{House prices}. OMI, a branch of the Italian Tax Office, twice per year   disseminates estimates of minimum and maximal home values in euros per square meter, $P_l$ and $P_h$, at very granular level. home values are available for all OMI micro-zones -- which are uniform socio-economic areas roughly corresponding to neighborhoods -- in Italian cities. $P_l$ and $P_h$ are estimated based on a limited sample of home sales and valuations by real estate experts. Further information is available at \url{https://www.agenziaentrate.gov.it/wps/content/Nsilib/Nsi/Schede/FabbricatiTerreni/omi}. 

We define the average home value in neighborhood (OMI micro-zone) $j$ as $\bar{P}_j = \frac{P_{lj}+P_{hj}}{2}$. The average home values at city level are estimated as a simple average of the $\bar{P}_j$. 
For further aggregation above the city level, we compute weighted averages of the cities' average home values. As weights, we use the stock of houses measured in the 2011 Census. OMI estimates are not designed for statistical purposes, and the index we construct must not be considered as equivalent to a quality-adjusted price index. 

In Italy, quality-adjusted (hedonic) house price indexes are disseminated by Istat, but their reference area is not consistent with our listing data, apart from three city-level indexes referred to the main Italian cities: Rome, Milan and Turin. We use these indexes for house prices in Rome and Milan in section \ref{sec:prices}.     
\vspace{0.5cm}

\noindent \textbf{Home sales}. Quarterly data about the volume of home sales in each city are disseminated by OMI.
\vspace{0.5cm}

\noindent \textbf{Italian Housing Market Survey}. The Italian Housing Market Survey is a quarterly survey that has been conducted by Banca d'Italia, OMI and Tecnoborsa since 2009. It covers a sample of real estate agents and reports their opinions regarding the current and expected course of home sales, price trends, time on market and terms of trade. See \url{https://www.bancaditalia.it/pubblicazioni/sondaggio-abitazioni/} for further information.
\vspace{0.5cm}

\noindent \textbf{Census data}. We retrieve detailed information on socio-economic characteristics and stock of buildings in OMI micro-zones from the 2011 Census. Istat census tracts are much smaller than OMI micro-zones (indeed, there are approximately 400,000 Istat census tracts over the Italian territory, as compared to 27,000 OMI micro-zones) and do not necessarily coincide with them. We perform spatial matching of the polygons representing the tracts and the micro-zones and impute the Istat variables to the OMI micro-zones according to the  overlap percentage of the polygons. For example, if an Istat census tract comprises 2,000 housing units and it straddles two OMI micro-zones, such that there is a 50\% overlap for both, we impute 1,000 housing units to each of the two OMI micro-zones. 

\section{Additional tables and figures}
\label{sec:addtabfig}

\begin{table}[ht]
\centering
\begin{tabular}{lcccccc}
  \hline
Variable & \multicolumn{5}{c}{Percentile} & Mean \\
\cline{2-6} 
         & 5 & 25 & 50 & 75 & 95 &   \\ 
  \hline
 Total population & 270.2 & 1604.5 & 4473.0 & 10592.2 & 27589.3 & 8049.8 \\ 
  Total households & 108.6 & 681.8 & 1874.5 & 4582.2 & 12226.4 & 3528.5 \\ 
  Total housing units & 152.6 & 896.5 & 2388.0 & 5527.2 & 13750.6 & 4166.2 \\ 
  Share of owner-occupied & 47.2 & 63.1 & 70.2 & 75.8 & 82.5 & 68.3 \\ 
  Average house value (\euro{}/$m^2$) & 765.6 & 1170.0 & 1532.5 & 2121.7 & 3339.0 & 1757.1 \\ 
  Average asking price (\euro{}/$m^2$) & 878.4 & 1290.1 & 1784.4 & 2453.8 & 4025.1 & 2013.5 \\ 
   \hline
\end{tabular}
\caption{Descriptive statistics of local markets.} 
\label{tab:descriptive_statistics_zoneomi}
\end{table}

\begin{figure}
\centering
\includegraphics[scale=0.45]{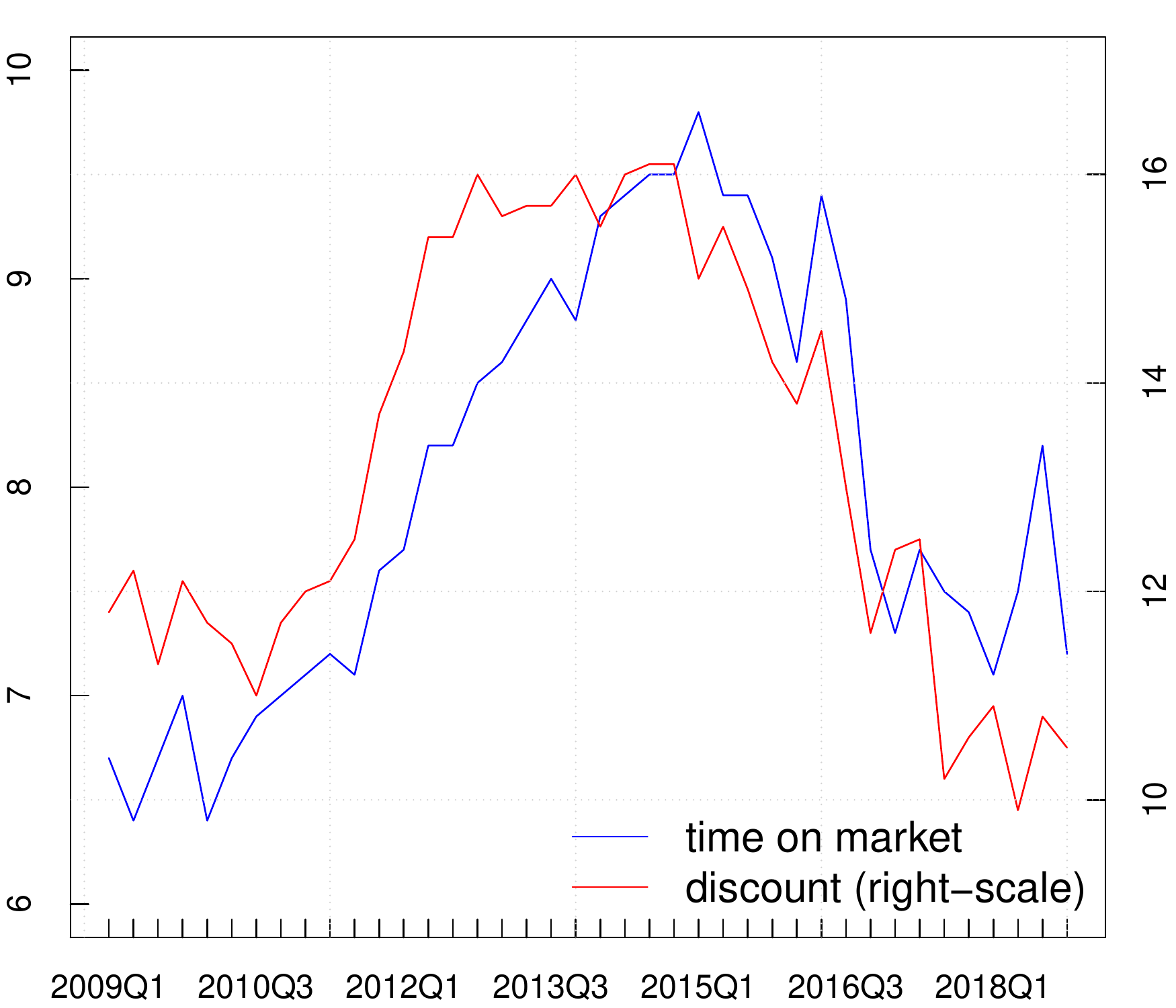}
\caption{Time on market and average discount in Italy.}
\label{fig:tom_discount}
\end{figure}

\FloatBarrier

\newpage

\section{Construction of the housing units dataset}
\label{sec:deduplication}

In this section we fully describe the algorithm we implemented to remove the duplicate ads. In \cite{loberto_etal_18} we also show the pseudo-codes of the procedure. 

\subsection{Pre-processing of the ads dataset}

We want to use the description of the housing unit to identify potential duplicates, but we first need to transform the text into a numeric vector using semantic analysis. There exist standard algorithms in natural language processing that accomplish this task by considering the multiplicity of the words, such as bag-of-words \citep{harris1954distributional}, but we cannot use these algorithms here. Indeed, two different real estate agents can describe the same dwelling using different words or sentences and this makes standard measures of distance among texts useless. For this reason we resort to the recent \textit{Paragraph Vector} (or \textit{doc2vec}) algorithm proposed by \cite{le_mikolov_2014}, an algorithm based on neural networks that allows to represent a document by a $N$-dimensional vector taking into account both the order and the semantic of the words. 

We also convert the class of some variables to alleviate the issue of misreporting dwellings characteristics. Indeed, two different agents can report information partially different but not completely at odds regarding the characteristics of the same housing unit. Consider for example the case of maintenance status: one real estate agent can report that the dwelling must be completely renovated, while the other agent writes that only a partial renovation is necessary. However, it is not plausible that the second agent says that the housing unit is new. As maintenance status takes only 4 possible ordered categories, we convert the categorical variable to an integer variable that takes value from 1 to 4 and a greater value means a better maintenance status. In this way when we compare two dwellings we take the absolute difference between the two variables and we can easily allow for partial matching. We do this operation for several other ordered categorical variables other than maintenance status: energy class, garage, type of garden, type of kitchen. We report the details in Table \ref{table:variables_trasformations}.   

\begin{table}[!h]
\centering
\footnotesize
\begin{tabular}{|l|L{4cm}|L{7cm}|}
  \hline
Variable & Original levels  & Transformation \\ 
  \hline
\textit{Garage} & Missing, Single, Double  & Integer: Missing = 0, Single = 1, Double = 2   \\ 
  \hline
\textit{Garden} & Missing, Shared, Private  &  Integer: Missing = 0, Shared  = 1, Private = 2    \\ 
  \hline
\textit{Maintenance status} & To renovate, Good, Excellent, New  & Integer:  To renovate = 0, Good = 1, Excellent = 2, New = 3   \\ 
  \hline
\textit{Kitchen Type}  & Kitchenette, Small eat-in kitchen, Large eat-in kitchen & Integer:  Kitchenette = 0, Small eat-in kitchen = 1, Large eat-in kitchen = 2  \\ 
  \hline
\textit{Energy Class} & A+, A, B, C, D, E, F, G  & Integer:  A+ = 0, A = 0, B = 1, C = 2, D = 3, E = 4, F = 5, G = 6    \\ 
  \hline
\textit{address} & Text of the address & Vector of words in the address (removing prepositions and articles)  \\ 
\hline
\end{tabular}
\caption{Variable transformations for the deduplication algorithm (classification tree).}
\label{table:variables_trasformations}
\end{table}

\subsection{Identification of duplicates}

We identify duplicate ads based on pairwise comparison, meaning that we compare each ad with all other ads that are potentially duplicates. 

First of all, in order to reduce the computational complexity of the pairwise approach, for each ad  we identify its potential duplicates. We define as potential duplicates those ads that refer to dwellings distant less than 400 meters and with a difference in asking price lower than 25\% in absolute value.\footnote{The difference in asking price is computed dividing the absolute difference between the two asking prices with the lowest of the two. Since this condition can be quite restrictive when considering dwellings with low asking prices, we consider as potential duplicates also those ads with absolute difference lower than 50,000 euro.} In this way we end up with a long list of pairs of ads and for each of them we have to decide if they are duplicates.

We classify each pair of ads as duplicates (TRUE) or distinct housing units (FALSE) based on a supervised classification tree. The algorithm adopted here is the C5.0 classification tree proposed by \cite{Quinlan1993c45} (\url{http://www.rulequest.com/see5-info.html}). This algorithm handles autonomously missing data, is faster than similar algorithms and allows for boosting.

For each pair of ads we provide to the algorithm a vector of predictors (covariates in the jargon of machine learning) and based on this information the classification tree returns the probability that the two ads are duplicates. We consider a pair of ads to be duplicates if the estimated probability is greater than 0.5.

Among the predictors we consider: floor area, price, floor, energy class, garage, garden type, air conditioning, heating type, maintenance status, kitchen type, number of bathrooms, number of rooms, janitor, utility room, location, elevator, balcony and terrace. For continuous variables, such as price and floor area, we use both the percentage and the absolute difference; for geolocation, we take the distance in meters between the geographical coordinates of the two dwellings. For binary variables, such as elevator or basement, the predictor is a dummy variable, that takes value equal to 1 if both ads share the same characteristic. For discrete ordered multinomial variables (such as maintenance status) we consider instead different degrees of similarity, by taking the absolute difference between the two variables. 

We also use the distance between the textual description of the two ads as a predictor. For  this variable we consider two different measures, depending on whether the ads are posted by the same agency or not. In the first case we use the Levenshtein distance, otherwise we compute the cosine similarities between the vectors produced using the \textit{Paragraph Vector} algorithm. 

We implement two different C5.0 models, depending on whether the ads are posted by the same agency or not. This choice is motivated by the observation that, when an agency posts two ads for the same dwelling, the characteristics in the ads are almost equal. On the contrary, when the ads are posted by different agencies (or by a private user) sometimes you can tell they refer to the same dwelling only thanks to the pictures on the website. This means that duplicate ads are less similar if posted by different agencies than if created by the same agency. As a consequence, a unique model for both cases could lead to an excess of ads considered as duplicates among those published by the same agency.

C5.0 is a supervised method that requires an initial training sample of pairs of ads of which we know with certainty whether they are duplicates or not. We construct two different training samples, one for each model, by manually checking the ads on the website, in particular comparing the pictures. The training sample for the ads of different agencies is made of 8296 pairs of ads; among them 3711 are duplicates (true positive, TP).  The training sample for the ads of the same agency is made of 9844 observations and 1850 are duplicates. These samples are constructed by iterating the following steps: (i) estimation of the model based on the initial training sample; (ii) out-of-sample validation of the models;  (iii) using the results of the out-of-sample exercise to increase the training sample. This three step approach is repeated several times, until we reach a sufficiently low misclassification error.  


\begin{table}
\centering 
\footnotesize
\begin{tabular}{L{5cm}ccccc}
\hline
&Observations & Duplicates & Precision & Recall & F-measure \\
\hline
Different agency  & 8296 & 3711 &   0.930 &  0.924  &  0.927 \\
Same agency & 9844 & 1850 &  0.952 &  0.946  & 0.949 \\
\hline
\end{tabular}
\footnotesize{ Precision = TP/(TP+FP).  Recall = TP/(TP+FN). F-measure = 2*(Precision*Recall)/(Precision+Recall). TP = true positive; FP = false positive; FN = false negative.} \normalsize
\caption{Assessment of C5.0 models}
\label{table:results_models}
\end{table}

In order to assess the performance of the two models we randomly split each training sample in two different sub-samples: the first one (90\% of the observations) is used to estimate the models, the second one (10\% of the observations) is used for the out-of-sample assessment of the classification performance. We repeat the operation 1,000 times and we evaluate the performance based on average results. Since the number of true negatives (ads that are not duplicates) is much larger than the number of true positives, using the classic accuracy rate can be misleading about the actual performance of the models. For this reason we consider measures of classification performance that do not rely on the number of true negatives, namely: precision, recall and F-measure.\footnote{The precision rate is defined as the ratio between the number of true positives and the sum of true and false positives; it thus measures how precise a classifier is in classifying true matches. The recall rate is defined as the ratio of true positives over the sum of true positives and false negatives; it measures the proportion of true matches that have been classified correctly. As there is a trade-off between precision and recall, we consider also a third additional measure, the F-measure, that calculates the harmonic mean between precision and recall.} 

We show the results in Table \ref{table:results_models}. As expected, the model for ads of the same agency is more precise than the one for ads of different agencies. As we said before, ads posted from the same agency and related to the same dwelling have almost all characteristics in common, therefore it is easier to identify them. However, as the F-measure is equal to .927, also the C5.0 model for ads of different agencies has a quite good classification performance. We should remark that the variables used in the two models are not the same and have been selected in order to maximize the  F-measure.\footnote{We started for both models with only five predictors: percentage difference between prices, absolute difference between prices, percentage difference between floor areas, absolute difference between floor areas, difference between floors. Then we added each candidate predictor one-by-one updating the initial model only if the variable provided an improvement of the F-measure (computed on the out-of-sample observations in a Monte Carlo experiment with 1,000 draws). We repeated the operation iteratively as long as there was no performance improvement from adding an additional predictor.} We report the set of variables for each model in Table \ref{table:variables_C50}.

\begin{table}[!ht]
\centering
\footnotesize
\begin{tabular}{|c|cc|L{9cm}|}
  \hline
Variable & Model 1 & Model 2 & Description of the variable \\ 
  \hline
\textit{price\_abs} & Yes & Yes & Absolute difference between asking prices \\ 
\textit{price\_per} & Yes & Yes & Percentage difference between asking prices   \\ 
\textit{floorarea\_abs} & Yes & Yes & Absolute difference between floor area  \\ 
\textit{floorarea\_per} & Yes & Yes & Percentage difference between floor area  \\   
\textit{floor} & Yes & Yes & Absolute difference between floor level (integer)   \\
\textit{distance} & Yes & Yes & Absolute distance in meters between households   \\   
\textit{address} & Yes & Yes & Indicator function: $1$ if the two addresses have at least one common word  \\ 
\textit{isnew} & Yes & Yes & Indicator function: $1$ if at least one of the ads refers to a new house \\ 
\textit{balcony} & Yes & No & Indicator function: $1$ if the feature balcony is the same\\ 
\textit{distdays1} & Yes & Yes & Number of days between the dates the ads have been added \\ 
\textit{status} & Yes & Yes & Absolute difference (integer) between categories   \\ 
\textit{elevator} & Yes & No & Indicator function: $1$ if the feature elevator is the same \\ 
\textit{energy\_class} & Yes & No & Absolute difference (integer) between categories   \\ 
\textit{isdetached} & Yes & No & Indicator function: $1$ if at least one of the ads refers to a detached or semi-detached house \\ 
\textit{bathrooms} & Yes & No & Absolute difference between number of bathrooms (integer)  \\ 
\textit{heating\_type} & Yes & No & Indicator function: $1$ if the feature heating type is the same  \\
\textit{distcontent1} & Yes & No & Cosine distance of vectors (\textit{Paragraph vectors}) representing textual descriptions  \\   
\textit{distcontent2} & No & Yes & Levenshtein distance between  textual descriptions  \\   
\textit{rooms} & Yes & No & Absolute difference between number of rooms (integer) \\ 
\textit{garage} & Yes & Yes & Absolute difference (integer) between categories  \\ 
\textit{garden} & Yes & No & Absolute difference (integer) between categories  \\ 
\textit{utility\_room} & Yes & No & Indicator function: $1$ if the feature utility room is the same \\ 
\textit{janitor} & Yes & No & Indicator function: $1$ if the feature janitor is the same \\
\textit{basement} & Yes & No & Indicator function: $1$ if the basement janitor is the same \\
\textit{is\_detached} & Yes & No & Indicator function: $1$ if the basement janitor is the same \\
\textit{pricemq\_abs} & No & Yes & Absolute difference in the asking price per square meter \\
\textit{pricemq\_min} & Yes & Yes & Minimum of the two asking prices per square meter \\ 
\textit{pricemq\_max} & No & Yes & Maximum of the two asking prices per square meter \\  
\hline
\end{tabular}
\caption{Variables for the classification trees}
\label{table:variables_C50}
\end{table}

\subsection{Creation of clusters of duplicates and information aggregation}
\label{sec:create_clusters}

Once we have identified the pairs of ads that are duplicates, we need a procedure to cluster all ads that are considered related to the same housing unit and to aggregate the information in the ads. Here, we follow a standard procedure in the computer science literature \citep{naumann_herschel_2010,christen_2012}

Let us suppose for example that we have only three ads: A, B and C. It is possible that the pairs (A,B) and (B,C) are considered as duplicates, but (A,C) is not. How should we manage this case? A simple solution is to assume transitivity: this means that since A is a duplicate of B and B is a duplicate of C, we assume that C is a duplicate of A and all these ads are considered related to the same dwelling. However, this approach can bring several issues: let us suppose for example that the probability of being duplicates for the pair (A,B) is 0.95 and the probability for the pair (B,C) is 0.51. How reliable is in this case the assumption of transitivity?

Here we abstract from the assumption of transitivity and we decide whether a cluster of ads refers to the same housing unit based on a measure of internal similarity of the cluster. In order to illustrate our approach we consider a simple example. Assume we have ten ads, we compute for each of the $45$ possible pairs the probability that they are duplicates and we remove all pairs with probability smaller than 0.5. The remaining pairs are shown in Table  \ref{table:example_graph}. 


\begin{table}[!h]
\centering
\footnotesize
\begin{tabular}{l|cccccccccc}
  \hline
 id.x & 1 & 1 & 2 & 4 & 4 & 4 & 6 & 6 & 7 & 9 \\
 id.y & 7 & 8 & 3 & 6 & 10 & 5 & 9 & 10 & 8 & 10 \\ 
 Prob. & 0.92 & 0.81 & 0.73 & 0.98 & 1.00 & 0.52 & 0.87 & 0.70 & 0.93 & 0.86 \\
   \hline
\end{tabular}
\caption{Example of clusters}
\label{table:example_graph}
\end{table}

\begin{figure}[!h]
\subfigure[Step 1] 
{\includegraphics[width=0.4\linewidth]{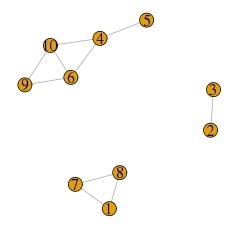} \label{fig:graph_1}}
 \hspace{15mm}
\subfigure[Step 2] 
{\includegraphics[width=0.4\linewidth]{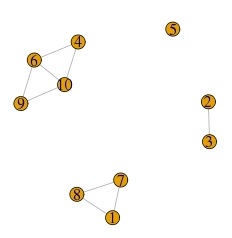}  \label{fig:graph_2}}
\caption{Clustering of the ads}
\end{figure}

Starting from the results of the pairwise classification step in Table \ref{table:example_graph}, we represent the information as a graph, in order to form clusters. The output of this step is represented in Figure  \ref{fig:graph_1}. The identifiers of the ads (here assumed to be integers between 1 and 10) are the nodes of the graph. Two nodes are connected if the probability that they are duplicates is greater than 0.5.

The tuples of ads (2,3) and (1,7,8) are considered to refer to two distinct dwellings, as each of the ads in the tuple is a duplicate of all the other ads. The troubles come with the tuple (4,5,6,9,10). Here, differently than before, it is not true that each ad is a duplicate of all the others. In particular this sub-graph only has 6 edges, while in order to be defined as a fully connected graph we would need 10 edges. More generally, an indirect graph is said to be fully connected if the number of edges is equal to $\frac{N(N-1)}{2}$, where $N$ is the number of the nodes of the graph (in our case the number of ads). 

The tuples (2,3) and (1,7,8) are clearly fully connected, while the tuple (4,5,6,9,10) is not. We consider a cluster as representing a single housing unit if it is a group of ads with a sufficiently high internal similarity, i.e. the number of edges is at least a fraction $5/6$ of the maximum number of edges that we can have in the cluster.
At each step we verify for each cluster if this condition is verified or not; if it is not satisfied we remove the weakest edge, that we define as the one with the lowest duplicate probability among those in the cluster.

Since for the tuple (4,5,6,9,10) the condition is not satisfied, we delete the weakest link, that in this case is represented by the edge between nodes (4) and (5), whose associated probability is 0.52. The new set of clusters after this operation is represented in Figure \ref{fig:graph_2}, in which the node (5) is now considered as referring to a distinct housing unit. If we look at the new tuple (4,6,9,10), we see that it has 5 edges out of 6 possible edges. Since our internal similarity condition is satisfied, we consider also this last tuple as a distinct dwelling.

Summing up our example, we started with 10 ads and we ended up with only 4 housing units. 
Based on this approach, we estimate for our training week (ads visible in 21 December 2016) that real dwellings were only 78\% of the total ads (130 thousands housing units out of 168 thousands ads).

Once we have created the clusters of ads identifying different dwellings, an additional issue that must be considered is to combine the information contained in multiple ads related to the same dwelling. Here, we adopt as a general rule that for each characteristic we take the one with highest absolute frequency. We deviate from this rule in the case of latitude and longitude (we compute the mean across the coordinates of all ads) and when we compute the dates of entry and exit of the dwelling into the housing market (for the entry we take the date of creation of the first ad associated to the dwelling, for the exit we consider the date of removal from the database of the last ad).\footnote{An additional exception to the general rule is done for asking prices. In this case we take the most frequent observation only among ads that have not been removed.} 

\subsection{Time machine approach}
\label{sec:timemachine}

The approach delineated above has the limit to be computationally unfeasible once the number of ads rises, because the number of pairwise comparisons increases exponentially. For this reason the procedure described in the previous section will be applied using an iterative approach (``time machine approach''). 

We process the ads progressively as soon as they are published on the website. At the first iteration of the process we run the deduplication procedure on all ads that have been added before the end of the first week we are considering. Once we apply the deduplication procedure, we end up with a new dataset where each row corresponds in principle to a unique dwelling and the characteristics of these housing units are derived from those of the associated ads.

At the second iteration we take as an input the datasets of ads and housing units of the first week. We check for duplicates only among the new ads added during the second week or the ads posted before but for which the price or other characteristics have been updated during the second week. For all these ads we look for duplicates both among new or updated ads and the dataset of housing units from the first week. The ads that are updated are preliminarily removed from the dataset of dwellings (that must be updated accordingly). 

The decision on whether the ads are duplicates is still based on a pairwise comparison, but now we can have pairs with two ads or pairs with one ad and one housing unit. Once we compute for each pair the probability that they are duplicates we cluster the results as explained in Section \ref{sec:create_clusters}. Differently than before, we impose the additional condition that in each cluster there can be at most one housing unit that was already identified in the previous week. This additional condition is necessary to avoid that clusters of ads that have been considered as referring to different dwellings in the past processing can be considered now as duplicates, because there are new ads that are potential duplicates of both of them.

\subsection{Additional controls}

After the deduplication process we make additional controls on the dataset to address for potential errors in the data. First of all we keep only the dwellings that have been on the market for at least two weeks. Then, we drop from the dataset those dwellings for which the price is not sufficiently consistent with the characteristics of the housing units. In this way we are also able to identify foreclosure auctions that were not previously identified, because for example the auction was not reported in the textual description . 

Our approach consists of running a hedonic regression, estimating for each dwelling the ratio between actual and predicted price and eliminating the housing units with a ratio between asking and predicted price lower than 0.5 or higher than 1.5.\footnote{We keep a relatively large range because the hedonic regression is limited to a small set of housing unit characteristics, those less affected by missing data issues. In this step we impute missing characteristics for each housing unit using the approach proposed by Honaker, King and  Blackwell (\url{https://gking.harvard.edu/amelia}).}

\end{document}